\PassOptionsToPackage{pdfpagelabels=false}{hyperref}
\documentclass[useAMS,fleqn,usenatbib]{mnras}
\pdfoutput=1 %
\usepackage{graphicx}
\usepackage{amsmath,amssymb,amstext}
\usepackage[T1]{fontenc}
\usepackage{ae,aecompl}
\usepackage[utf8]{inputenc}
\usepackage{newtxtext,newtxmath}
\usepackage[figure,figure*]{hypcap}
\usepackage[dvipsnames]{xcolor} 
\usepackage{bm}
\usepackage[normalem]{ulem}

\usepackage{xparse}

\NewDocumentCommand{\INTERVALINNARDS}{ m m }{
    #1 {,} #2
}
\NewDocumentCommand{\interval}{ s m >{\SplitArgument{1}{,}}m m o }{
    \IfBooleanTF{#1}{
        {#2 \INTERVALINNARDS #3} #4
    }{
        \IfValueTF{#5}{
            #5{#2} \INTERVALINNARDS #3 #5{#4}
        }{
            #2 \INTERVALINNARDS #3 #4
        }
    }
}

\input{macros.sty}
\input{hyperlink-year-only-natbib-patch}

\title[Constraining Assembly Bias]{How to Optimally Constrain Galaxy Assembly Bias: Supplement Projected Correlation Functions with Count-in-cells Statistics}

\author[K.~Wang et al.]{%
Kuan~Wang,$^{1,2}$\thanks{E-mail: \email{kuw8@pitt.edu}},
Yao-Yuan~Mao$^{1,2}$\thanks{E-mail: \email{yymao.astro@gmail.com}},
Andrew~R.~Zentner$^{1,2}$, 
Frank~C.~van~den~Bosch$^{3}$, \newauthor
Johannes~U.~Lange$^{3}$, 
Chad~M.~Schafer$^{4,5}$, 
Antonio~S.~Villarreal$^{1,6}$, \newauthor
Andrew~P.~Hearin$^{6}$, and 
Duncan~Campbell$^{5,7}$
\vspace*{6pt}
\\
$^{1}$Department of Physics and Astronomy, University of Pittsburgh, Pittsburgh, PA 15260, USA\\
$^{2}$Pittsburgh Particle Physics, Astrophysics, and Cosmology Center (PITT PACC), University of Pittsburgh, Pittsburgh, PA 15260, USA\\
$^{3}$Department of Astronomy, Yale University, P.O. Box 208101, New Haven, CT 06511, USA\\
$^{4}$Department of Statistics \& Data Science, Carnegie Mellon University, Pittsburgh, PA 15213, USA\\
$^{5}$McWilliams Center for Cosmology, Carnegie Mellon University, Pittsburgh, PA 15213, USA\\
$^{6}$Argonne National Laboratory, Argonne, IL 60539, USA\\
$^{7}$Department of Physics, Carnegie Mellon University, Pittsburgh, PA 15213, USA
}
\date{22 March 2019}

\pubyear{2019}

\begin{document}
\label{firstpage}
\pagerange{\pageref{firstpage}--\pageref{lastpage}}
\maketitle

\begin{abstract}

Most models for the statistical connection between galaxies and their haloes ignore the possibility that galaxy properties may be correlated with halo properties other than halo mass, a phenomenon known as galaxy assembly bias. And yet, it is known that such correlations can lead to systematic errors in the interpretation of survey data that are analyzed using traditional halo occupation models. At present, the degree to which galaxy assembly bias may be present in the real Universe, and the best strategies for constraining it remain uncertain. We study the ability of several observables to constrain galaxy assembly bias from redshift survey data using the decorated halo occupation distribution (dHOD), an empirical model of the galaxy--halo connection that incorporates assembly bias. We cover an expansive set of observables, including the projected two-point correlation function $\wprp$, the galaxy--galaxy lensing signal $\Dsigma$, the void probability function $\VPF$, 
the distributions of counts-in-cylinders $\Pncic$, and counts-in-annuli $\Pncia$,
and the distribution of the ratio of counts in cylinders of different sizes $\Pratio$. We find that despite the frequent use of the combination $\wprp+\Dsigma$ in interpreting galaxy data, the count statistics, $\Pncic$ and $\Pncia$, are generally more efficient in constraining galaxy assembly bias when combined with $\wprp$. Constraints based upon $\wprp$ and $\Dsigma$ share common degeneracy directions in the parameter space, while combinations of $\wprp$ with the count statistics are more complementary. Therefore, we strongly suggest that count statistics should be used to complement the canonical observables in future studies of the galaxy--halo connection.

\end{abstract}

\begin{keywords}
cosmology: large-scale structure of Universe -- galaxies: evolution -- galaxies: formation -- galaxies: haloes -- galaxies: statistics
\end{keywords}

\section{Introduction}
\label{sec:intro}

In the concordance $\Lambda$CDM model of the Universe 
\citep{komatsu_WMAP711,2014AAS...22342704B,planck13,Planck2018,DES2018}, 
galaxies reside in dark matter haloes \citep{whiterees78,blumenthal_etal84}, 
which form around peaks in the primordial dark matter density field 
\citep{bardeen_etal86,Bond91,sheth_tormen01,sheth01,zentner07}. 
In practice, the abundance, clustering, and structure of dark matter 
haloes have been precisely documented by high-resolution, gravity-only 
$N$-body simulations of cosmological structure growth 
\citep[e.g.,][]{nfw97,mo_white96,tinker08a,tinker_etal10}. 
Halo occupation models use empirical data to link galaxies to haloes in a statistical sense. 
They are useful 
because they provide a convenient means to compare observed galaxy 
clustering statistics with theoretical predictions without a complete 
theory of galaxy formation and evolution. This, in turn, is useful because 
one can use such models to test cosmological models using data on non-linear 
scales, and because empirical models distil the formidable amount of 
information available in survey data into a relatively simpler 
galaxy--halo relationship that can be used to inform models of galaxy 
formation and evolution. \citet{wechsler_tinker18} provide a contemporary 
review of these models.
 
The term {\em assembly bias} has, unfortunately, taken on several related
but distinct meanings in the literature. The clustering of dark matter 
haloes is a strong function of halo mass \citep{kaiser84,mo_white96}, 
but it has become clear over the last decade that haloes cluster 
as a function of a number of other properties 
\new{\citep{gao_etal05,wechsler06,gao_white07,li_etal08,mao_etal18, Xu_2018}}. \citet{gao_etal05}
first studied the age-dependence of halo clustering, which led to the 
term {\em assembly bias}, but because many halo properties are correlated 
with formation history, the dependence of halo clustering on many other 
properties (e.g., concentration, spin, and so on) is often loosely 
referred to as {\em assembly bias} or {\em halo assembly bias} as well. 
\citet{mao_etal18} advocate referring to these dependences as 
{\em secondary biases}\footnote{The {\em primary bias} is the 
strong dependence of halo clustering on halo mass.}. 
This nomenclature is clearer because the secondary biases 
do not necessarily have a clear origin in the correlations 
of halo properties, such as concentration, with conventional
measures of halo formation history.

\new{Assembly bias of observed galaxies as well as simulated halo 
populations has received significant attention in the recent literature regarding
the analysis and interpretation of galaxy survey data. Assembly bias 
may challenge survey analyses because it may induce (1) systematic errors in 
the inferred galaxy--halo relationship inferred from survey data 
\citep[e.g.,][]{zentner_etal14} and/or (2) biases in inferred cosmological 
parameters\citep[e.g.,][]{croton_etal07,wu08, McCarthy_2018}. On the other hand, an unambiguous detection of assembly bias in 
survey data may pave the way to a richer and more complete understanding 
of the connection between galaxies and their host dark matter haloes.}

In the context of the interpretation of galaxy surveys, assembly bias 
can lead to the following possibility. Galaxies in a certain luminosity sample may form in haloes 
with a probability that depends upon not only the mass of the halo, but on any number of halo properties. If this happens, then the resultant 
clustering of any galaxy sample must be interpreted within the context of 
a model that incorporates not only the mass 
dependence of halo clustering, but also the secondary biases, \new{as is done in e.g., \citet{chaves-montero_2016, zentner_etal16, Lehmann2017, Romano-Diaz_2017, desjacques_2018}}. 
This case is sometimes loosely referred to as {\em galaxy assembly 
bias.} \new{Signals of galaxy assembly bias are also present in hydrodynamical simulations \citep[e.g.,][]{xu_zheng_2018_illustris, Artale_2018, Bose_2019}.} The issue facing survey data analysis is that the vast majority 
of studies treat survey data using models that assume that galaxies of 
a particular luminosity (or other galaxy properties that determine 
the sample selection) form within haloes with a probability that depends 
upon only the mass of the halo (and no other halo property). Thus, these analyses 
account only for the mass dependence of halo clustering 
\citep{wechsler_tinker18}.

The most widely used empirical models for interpreting survey data \new{include} the 
halo occupation distribution (HOD) \citep[e.g.,][]{berlind02} and the 
conditional luminosity function (CLF) \citep[e.g.][]{yang03}. 
Both of these models in the original (and standard) 
forms assume that galaxies of a particular type reside in haloes with 
a probability that depends only on the masses of the haloes. 
Both the HOD 
\citep{zheng07, zehavi_etal11, guo_etal12, guo_zehavi_zheng12, leauthaud_etal12, guo_etal14, zentner_etal16, vakili_etal16}
and the CLF 
\citep{vdBosch03a, yang_etal04, cooray06, vdBosch07, cacciato_etal13,lange2018}
have been used successfully to interpret a variety of 
observational samples. 
\new{On the other hand, subhalo abundance matching \citep[AM or SHAM, e.g.,][]{kravtsov04a, tasitsiomi04, vale_ostriker04, trujillo_gomez11, hearin_etal13a} has the power to naturally incorporate assembly bias, by matching galaxy properties to halo properties (e.g., $V_{\rm{peak}}$) that can assume varying values at a fixed halo mass. The success of SHAM in interpreting galaxy survey data \citep{behroozi10, behroozi13, Reddick2013, Lehmann2017} is striking considering the simplicity of its assumptions.}

In this paper, we will use an expanded form of the HOD, known as the 
decorated HOD \citep[dHOD,][]{hearin_etal16}. The dHOD builds upon the traditional HOD by adding 
parameters that enable tunable levels of galaxy assembly bias. 
We limit our treatment to the HOD and dHOD for 
specificity and simplicity.

We explore the utility of several galaxy survey observables to constrain 
assembly bias within the context of simple dHOD \citep{hearin_etal16} models. 
Aside from the overall galaxy number density $\ngal$, which we use as a basic constraining observable throughout our analyses, the observables that we explore are the projected galaxy correlation function 
$\wprp$, the excess surface density inferred from galaxy--galaxy lensing $\Dsigma$, the void probability function 
$\VPF$, galaxy counts-in-cylinders $\Pncic$, galaxy counts-in-annuli $\Pncia$, and the probability distribution of the ratio between $\Ncic$ of different cylinder sizes $\Pratio$. In this first study of the subject, we intentionally avoid utilizing 
satellite kinematics (SK), redshift space distortions (RSD), or  
other observables that require a detailed model of galaxy velocities relative 
to haloes. Treating such statistics requires additional modelling and additional 
assumptions that can greatly complicate such a study. \new{For the observables we study, we examine and compare their effectiveness at constraining not only assembly bias 
within the context of a dHOD model, but nearly all HOD parameters.}

The remainder of this paper gives the details necessary to support the 
summary of our findings stated in the previous paragraph. In 
Section~\ref{sec:methods}, we describe the simulation that we use, the 
dHOD models within which we work, the observables we consider, 
and our approach to estimating parameter constraints. 
In Section~\ref{sec:results}, we present our results in detail. 
We discuss our results in the context of the contemporary literature, 
draw broad conclusions, and propose future steps in Section~\ref{sec:discussion}.

\section{Methods}
\label{sec:methods}

In this section, we give the details of our analysis procedures. This 
includes a discussion of the simulation that we use, the dHOD models 
that we explore, the observables that we consider, our methods for 
estimating parameter constraints, and our estimates of the 
covariance matrices used in our analyses.

\subsection{Simulation}
\label{sec:simulation}

In order to mitigate the limitations of analytic estimates of clustering 
and lensing statistics \citep[see e.g.,][]{vdbosch13}, the calculations that we perform in this paper are 
based upon $N$-body simulations of the formation of structure in a concordance 
cosmological model. These simulations
evolve dark matter particles under the influence of gravity from 
initial over-densities in the early universe to the present day. 
In particular, the analysis in this work utilizes the Bolshoi Planck 
simulation\footnote{\new{Available at }\https{www.cosmosim.org/cms/simulations/bolshoip}}
\citep{klypin_etal16,rockstar,behroozi_trees13,bolplanck2016}. 
Bolshoi Planck  is a dark matter only simulation 
within a cubic box of length $250\, \Mpch$, 
which adopts values of cosmological parameters from \citet{planck13}, 
namely $\Omega_{\Lambda}$ = 0.693, 
$\Omega_\mathrm{m}$ = $1-\Omega_{\Lambda}$ = 0.307, 
$\Omega_\mathrm{b}$ = 0.048, h = 0.7, 
$n_\mathrm{s}$ = 0.96, and $\sigma_8$ = 0.82. 
The simulation contains 2048$^3$ particles, 
implying a particle mass of 
$m_{\mathrm{p}} = 1.55 \times 10{^8}\, \Msunh$.

We use the \texttt{bolplanck} halo catalogue 
included with the \texttt{Halotools} 
software package\footnote{\https{halotools.readthedocs.io}}
\citep{halotools}, 
which also provides an implementation of customizable dHOD models.

The \texttt{bolplanck} halo catalogue was 
produced from the Bolshoi Planck simulation using the \texttt{ROCKSTAR} 
halo-finder \citep{rockstar}. To compute lensing observables, we 
use the particle catalogue included with \texttt{Halotools} which contains 
$10^6$ randomly-selected particles from the Bolshoi Planck volume, and make a 
downsampled catalogue containing $\sim10^5$ particles with an 
acceptance rate of 0.1 for runtime considerations. We have tested that the measured $\Dsigma$ is not sensitive to the downsampling, and the noise introduced in this process is accounted for in our covariance matrix (see Section~\ref{sec:cov}).
The catalogues are included in 
\texttt{Halotools} version \texttt{halotools\_v0p4}, which adopts the virial definition of haloes, and we work at $z=0$, corresponding to an overdensity parameter $\Delta_{\textrm{vir}}=333$ with respect to the mean matter density of the Universe.

\subsection{Halo Occupation Model}
\label{sec:hod}

We describe the galaxy--halo connection using the HOD and the dHOD. 
Both of these models specify the probability for a halo of mass 
$\Mvir$ to host $\Ncen$ central galaxies and $\Nsat$ satellite 
galaxies above a certain threshold stellar mass, $P( \Ncen \vert \Mvir)$ and $P( \Nsat \vert \Mvir)$ 
respectively. Central and satellite galaxies are considered separately 
because central galaxies reside in the potential wells of host haloes while satellite 
galaxies are associated with subhaloes and experience different physics of formation and evolution. 
It is well known that subhaloes experience very distinct evolution from host haloes and 
thus have demographics that are distinct from host haloes \citep{kravtsov04a,zheng05,zentner_etal05}. 
Moreover, numerous observations, using many different approaches, have established that 
central galaxies and satellites have different properties and can be described as 
two distinct populations \citep{oemler74,dressler80,postman_geller84,zehavi05a,weinmann06b,vdbosch08}.

The central galaxy occupation is modelled as a Bernoulli \new{random variable, which takes the value 1 with probability $p$ and the value 0 with probability $1-p$, with $p=\langle \Ncen \vert \Mvir \rangle$}. Satellite galaxies follow a Poisson distribution 
with first moment $\langle \Nsat \vert \Mvir \rangle$. 

In principle, the formalism we use in this work is the
same as in \citet{zentner_etal16} which, in turn, 
was chosen to mimic the analysis of \citet{zehavi_etal11}. 
However, we adopt the implementation of the model in \texttt{Halotools}, and introduce some subtle modifications which will be elaborated on in Section~\ref{sec:decorated}.

\subsubsection{Standard HOD}
\label{sec:standard}

In the standard HOD, which does not account for any potential galaxy assembly 
bias, the mass of a halo solely determines the 
galaxy occupation. The mean central and satellite galaxy 
occupations vary with halo mass according to 
\begin{eqnarray}
\langle \Ncen \vert \Mvir \rangle & = & \frac{1}{2} \left( 1 + 
\mathrm{erf} \left[\frac{\log(\Mvir)-\log(M_{\textrm{min}})}{\slogM}  \right]  \right),\\
\langle \Nsat \vert \Mvir  \rangle & = & \left(\frac{\Mvir - \Mzero}{\Mone}\right)^{\alpha} \times \langle \Ncen \vert \Mvir \rangle,
\label{eq:nstd}
\end{eqnarray}
\citep{zheng07}, where $\Mmin$ is the mass at which a halo has a 50\% 
probability of hosting a central galaxy; 
$\sigma_{\log M}$ is a measure for the scatter in the stellar mass--halo mass relation that determines the steepness of the $\langle \Ncen \vert \Mvir \rangle$ transition from zero 
to unity; $\Mzero$ is the 
truncating mass, below which $\langle \Nsat \vert \Mvir\rangle = 0$; 
the mass $\Mone$ indicates the halo mass at which there is, on average, one satellite\footnote{More accurately, this mass
is $\Mone$+$\Mzero$, but $\Mzero$ is typically much smaller than $\Mone$.} if a central is present;
and, finally, $\alpha$ is the index of the satellite occupation power law. 
Note that Eq.~(\ref{eq:nstd}\new{)} expresses the probability of having 
a satellite galaxy for a halo with mass $\Mvir$, after marginalizing 
over the central occupation. The first term on the right hand side indicates 
the mean satellite occupation in haloes with a central galaxy, while the
second term modulates this occupation by the probability for a halo to contain
such a central. Hence, the presence of a central boosts the probability 
for a halo to host satellite galaxies. Note, though, that for individual
haloes a central galaxy is not strictly required for satellites to be present.
Although this modulation with $\langle \Ncen \vert \Mvir \rangle$ 
is fairly common
\citep[e.g.,][]{zheng07, zehavi_etal11, zentner_etal16}, we emphasize that it is not used by all authors.

These specifications, along with the assumptions that 
the central galaxy HOD is a Bernoulli distribution 
and the satellite galaxy HOD is a Poisson distribution, 
suffice to specify fully the halo occupation statistics of dark matter haloes in a standard HOD model
without assembly bias.

\subsubsection{Decorated HOD}
\label{sec:decorated}

Galaxy assembly bias can be incorporated into the HOD formalism in any 
number of ways. For a secondary halo property $x$, 
(e.g., concentration, spin, etc.), one can specify a 
functional form for the probability distributions $P(\Ncen \vert \Mvir, x)$ 
and $P(\Nsat \vert \Mvir, x)$. In such a generalized HOD, the clustering 
of galaxies can be altered if halo clustering depends upon secondary 
property $x$. The decorated HOD \citep[dHOD,][]{hearin_etal16} is one way of 
incorporating assembly bias into the HOD formalism such that integrating the 
dHOD probability distributions over the secondary properties of interest 
yields the standard HOD.

In the present paper, we use a simple variation of the dHOD as an illustrative 
model. In particular, we divide haloes into two categories based upon 
secondary halo property $x$. Haloes with higher values of $x$ are 
assigned distinct HODs compared to haloes with lower values of $x$, 
with a pivot value of $x_{\mathrm{piv}}$. 
This is the ``discrete halo subpopulations'' example discussed in 
Section~4.2 of \citet{hearin_etal16} and used to analyze SDSS data \citep{DR7_09}
in \citet{zentner_etal16}. To specify completely the dHOD, 
we assume that $P(\Ncen \vert \Mvir, x)$ is a Bernoulli distribution and that 
$P(\Nsat \vert \Mvir, x)$ is a Poisson distribution, but that these 
distributions have first moments of 
\begin{eqnarray}
\langle \Ngal \vert \Mvir, x>x_{\mathrm{piv}} \rangle & = & 
\langle \Ngal \vert \Mvir \rangle + \delta \Ngal, \label{eq:ngaldec1}\\
\langle \Ngal \vert \Mvir, x \le x_{\mathrm{piv}} \rangle & = &
\langle \Ngal \vert \Mvir \rangle - \delta \Ngal,
\label{eq:ngaldec2}
\end{eqnarray}
where we use the notation $N_{\textrm{gal}}$ because this modification 
applies equally well to both the central and satellite occupations. 
We choose $x_{\mathrm{piv}}$ to be the median value of $x$ 
at a given halo mass, 
so that each population contains 50\% of all the haloes.
In this toy model, assembly bias manifests itself as a step function 
in the secondary property $x$, though we expect that any assembly 
bias realized in nature would be represented by a smooth function of $x$. 
This simple model is practical in the sense that current data are 
not sufficient to constrain more complex models \citep{zentner_etal16}; 
however, richer models of assembly bias can naturally be 
accommodated within the dHOD framework \citep{hearin_etal16} 
and future data sets are likely to enable constraints on richer models.

The differences $\delta \Ngal$ above are characterized by two assembly bias parameters, $\Acen$ and $\Asat$, both constructed so that they range between 
1 and -1, in addition to the five standard HOD parameters. A list of the 7 dHOD parameters can be found in Table~\ref{tab:fidp}. Positive values of $\Agal$ indicate a
positive correlation between galaxy number and 
halo property $x$ (i.e., haloes with $x>x_{\mathrm{piv}}$ 
contain more galaxies, on average, 
than those with $x<x_{\mathrm{piv}}$),
while negative values represent anti-correlation. 
When $\Agal=0$, the model reduces to the traditional standard HOD. 
Note that $\Acen$ and $\Asat$ vary independently of one another and do 
not necessarily have the same sign. The stipulations that the occupation 
of a halo never be negative, and the requirement that
\begin{equation}
  \langle \Ngal \vert \Mvir \rangle = \int \langle \Ngal \vert \Mvir, x\rangle \, P(x \vert \Mvir) \, {\rm d}x\,,
\end{equation}
with $P(x\vert\Mvir)$ the probability distribution for $x$ given $\Mvir$, implies that
\begin{eqnarray}
\delta \Ncen & = & \Acen \, \mathrm{min}
\left[\langle \Ncen \vert \Mvir \rangle, 1-\langle \Ncen \vert \Mvir \rangle\right],\\
\delta \Nsat & = & \Asat \, \langle \Nsat \vert \Mvir \rangle.
\label{eq:dngal}
\end{eqnarray}

It should be noted that, when populating a mock galaxy catalogue using HOD or dHOD, the actual number of galaxies in each halo is a random variable: the number of central galaxies follows the Bernoulli distribution, and the number of satellite galaxies follows the Poisson distribution. 
Since we will be conducting a Fisher analysis, the random fluctuation in realizations can masquerade as a dependence of galaxy number density on (d)HOD parameters, and yield artificially tight constraints. Hence, we need to reduce the random fluctuation in realizations as much as possible so that a small change in one or more (d)HOD parameters results 
in a small change in the total number of galaxies. 
We achieve this by assigning to each halo two random variates, 
$p_\text{cen}$ and $p_\text{sat}$, both drawn from the uniform 
distribution $U(0, 1)$, 
independently from the (d)HOD parameter values. 
We then find the number that corresponds to these $p$-values 
in the cumulative distribution of a Bernoulli distribution 
(for $p_\text{cen}$) or a Poisson distribution (for $p_\text{sat}$). 
This minimizes the random fluctuations among realizations that 
only differ slightly in their corresponding (d)HOD parameters.

In the case of the dHOD, the mean number density of galaxies is 
strictly independent of the dHOD parameters $\Acen$ and $\Asat$, 
and so the problem of preserving the total number density from one 
mock realization to another is particularly acute. 
In the dHOD, changes to $\Acen$ or $\Asat$ result in changes to 
the \emph{mean} occupations of individual haloes, but should 
result in no change to the total number density. 
If the galaxy occupation for each halo is realized 
independently, then the total number of galaxies can 
vary from mock realization to another as $\Acen$ and/or 
$\Asat$ are varied. 
The result of such a variation would be to infer additional 
constraining power on $\Acen$ and $\Asat$ where there should be none. 
To mitigate this possibility, we slightly 
modified the dHOD implementation in \texttt{Halotools}\footnote{Our implementation is called \texttt{PreservingNgalHeavisideAssembias} in \texttt{Halotools}.} 
to ensure the total number density of galaxies is preserved among 
mock catalogs that differ only in their values of $\Acen$ and 
$\Asat$. We achieved this by conditioning the dHOD on 
the total number of galaxies before realizing 
the occupation of each individual halo.
It should be noted that once the total number of 
central galaxies is fixed, the number of central galaxies 
in each halo would no longer be strictly a 
Bernoulli distribution. However, for satellite galaxies, both 
the total number of galaxies and the number of galaxies in an individual 
halo follow Poisson distributions.

In this work, we choose the NFW concentration parameter 
\citep{nfw97} as our 
secondary property 
\new{[so $x = c_{\rm{NFW}}$ in 
Eqs.~(\ref{eq:ngaldec1}) and (\ref{eq:ngaldec2})] 
when studying constraints on the parameters $\Acen$ and $\Asat$. 
As has been shown in \citet{croton_etal07} and 
\citet{villarreal2017}, concentration only partially accounts 
for galaxy assembly bias, and other halo properties 
(e.g., halo age, spin, environment density) may also contribute 
to assembly bias. 
Nevertheless, we choose concentration for several physically motivated reasons.}
First, concentration is known to correlate with assembly 
history \citep{wechsler02}\footnote{Though this does 
not guarantee that concentration and 
assembly history metrics will lead to similar secondary biases 
\citep{mao_etal18}.}, and has the advantage that 
it can be measured in a single snapshot of a 
simulation. \new{Second, the success of abundance matching suggests that 
the HODs realized by nature may, indeed, have some dependence upon halo 
structure \citep{conroy06,Reddick2013,HW13a,zentner_etal14,Mao2015}. 
Indeed, \citet{Lehmann2017} showed that abundance matching in a manner 
that does not include any concentration dependence is excluded by 
galaxy clustering.} 
Third, haloes are known to exhibit large concentration-dependent 
clustering in the mass range of interest to us 
($\sim$~a few $\times 10^{12}\ \Msunh$). \new{Consequently, 
concentration-dependent clustering is an excellent test case 
with which to study methods to constrain assembly bias.}
Fourth, concentration-dependent clustering has already been 
studied in \citet{hearin_etal16} for the dHOD and for 
observational samples by \citet{zentner_etal16} and 
\citet{vakili_etal16}, providing a baseline for comparison.
\new{Given the above reasons, we believe the concentration parameter 
is the most reasonable choice for this study, 
yet we note that our findings may not be 
trivially generalized to the assembly bias induced 
by other secondary halo properties, 
as other properties may induce different assembly bias behaviours.}

\subsubsection{Spatial and Velocity Distribution}
\label{sec:galprof}

The detailed predictions of an empirical model depend not only on the 
model for halo occupation, but also upon the positions and velocities, 
relative to the host halo, that are assigned to the galaxies. 
We place the central galaxy at the halo centre and the central galaxy inherits 
the host halo's peculiar velocity. 
Satellite galaxies are distributed within the virial radius of the 
host halo according to a spherically symmetric NFW profile 
characterized by the same concentration as the 
dark matter distribution. This assumption is supported by various works \citep[e.g.,][]{vdMarel.etal.00, lin_mohr_2004, vdB05a}, though other authors find that the 
distribution of satellite galaxies are described by a concentration different 
from that of dark matter particles, depending on the satellite population 
\citep{Chen2008,more09a,watson_etal12a,tal12, lange_etal18}.
The radial velocity distribution of satellite galaxies is modelled as a 
Gaussian distribution with the host halo velocity as the first moment 
and the solution of the isotropic Jeans equation for an 
NFW profile \citep{klypin99a} as the second moment.
We assume velocities to be isotropic, and draw the peculiar velocities in each
Cartesian direction independently from this distribution. In practice, the statistics that 
we examine are quite insensitive to moderate alterations to the treatment 
of galaxy peculiar velocities (this is by design), 
though it would be interesting to explore 
statistics that are sensitive to peculiar velocities as a follow-up 
study.
\new{To examine the effect of alternative velocity models, we have tested the velocity bias model in \citet{guo_2015} with $\alpha_c=0.3$ and $\alpha_s=1$, and find that the systematic change in our observables is negligible (within $1.5\%$ in all cases).}

\subsection{Observables}
\label{sec:obs}

In search of effective ways of utilising existing and future galaxy surveys to constrain the dHOD, we consider a number of observables that are sensitive to halo occupation. 
In particular, while including the overall galaxy number density of the simulation volume, $\ngal$, as a constraining observable in all of our analyses, we examine 
\begin{enumerate}
    \item the projected two-point correlation function, $\wprp$;\\
    \item the galaxy--galaxy lensing signal, $\Dsigma$;\\ 
    \item the void probability function $\VPF$;\\
    \item the distribution of counts-in-cylinders, $\Pncic$;\\
    \item the distribution of counts-in-annuli, $\Pncia$
(analogous to counts-in-cylinders, but with an excised inner region);\\
    \item the distribution of the ratio of counts in cylinders of different sizes $\Pratio$.
\end{enumerate}
We discuss these observables in more detail in 
the remainder of this subsection.

\begin{figure*}
\centering
  \includegraphics[width=\textwidth]{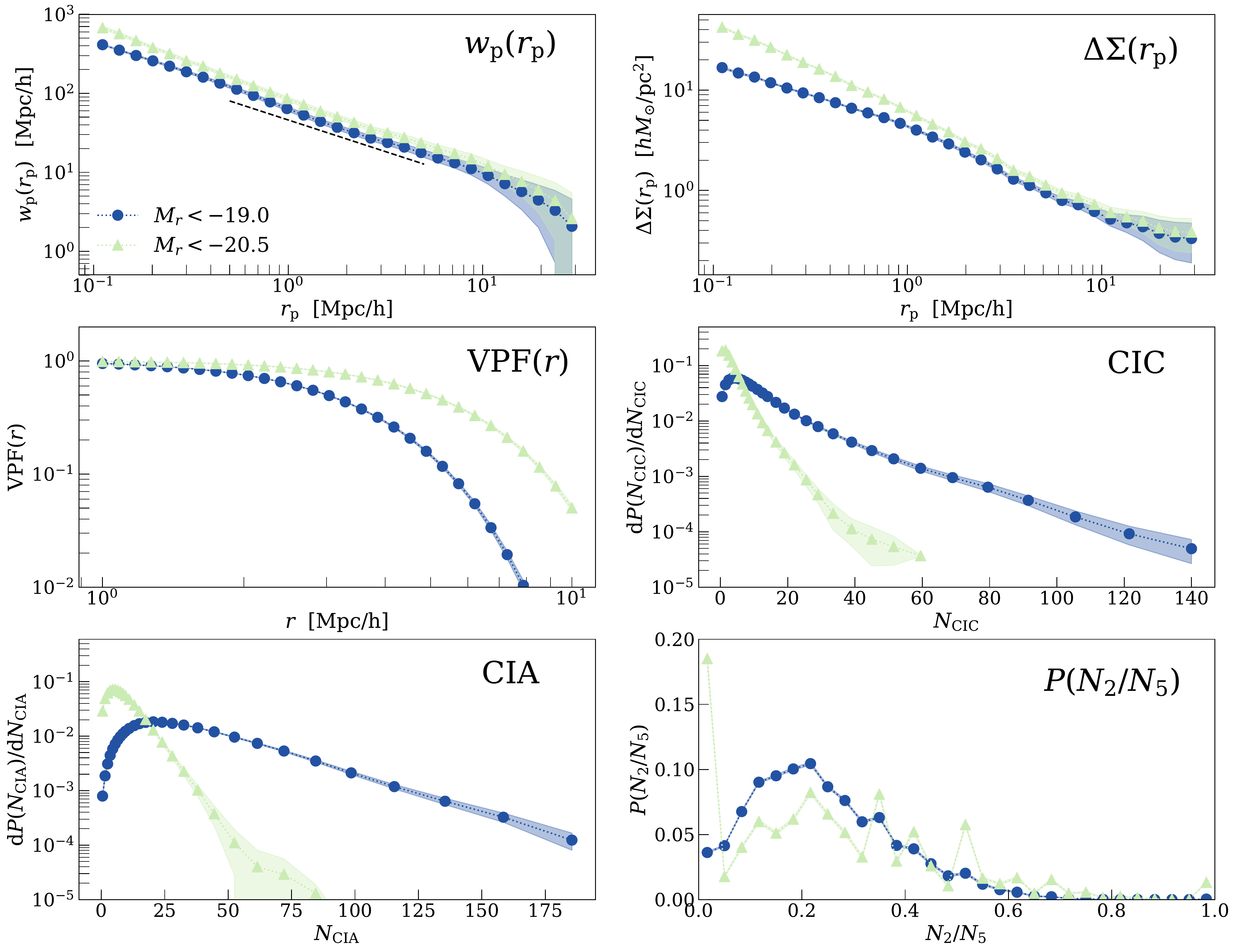}
  \caption{\label{fig:obs_example}
  Examples of the observables that we consider to constrain assembly 
  bias. In each panel, we show examples of the observables. We show examples for our fiducial HOD parameters corresponding to two luminosity thresholds, 
  as illustrated by the legend in the upper left panel. \new{The observable values are shown as connected data points, while the coloured bands show the uncertainty from jackknife subsampling, which we describe in detail in Section~\ref{sec:cov}.}
  The upper left panel shows the 
  projected correlation function, $\wprp$. For illustration, this panel 
  also contains a dashed line illustrating the slope of a power law with 
  $\wprp \propto \rp^{-0.8}$ for comparison. The upper right 
  panel shows the excess surface density about galaxies, $\Dsigma$, in the 
  samples. The left panel in the middle row depicts 
  the void probability function, $\VPF$. The right panel 
  in the middle row depicts the distribution of counts-in-cylinders 
  (CIC). Notice that the lower luminosity sample has a much more significant 
  tail to high companion counts than the higher luminosity sample. 
  Similarly, the left panel of the bottom row depicts counts-in-annuli 
  (CIA). Finally, the right panel of the bottom row shows 
  the probability distribution of the ratio of cylinder counts on distinct scales, 
  $\Pratio$. Each panel is labelled by the observable shown.}
\end{figure*}

We compute all observables numerically, by generating mock galaxy catalogues  
and subsequently measuring each observable from the mock catalogue. 
This forward-modelling approach enables us to mitigate modelling 
uncertainty associated with analytic approaches to galaxy clustering 
and to incorporate possible systematic errors into our calculations. 
All observables are computed in redshift space, as they would be 
from observational data, namely, 
the coordinates of galaxies $(x,y,z)$ are mapped onto $(x,y,z+v_z/aH(a))$. 
We show examples of the measured values of the observables \new{and their uncertainties from jackknife subsampling (see Section~\ref{sec:cov})} in 
Fig.~\ref{fig:obs_example} for our fiducial HOD models. 
Our fiducial models are taken from the fits of \citet{zentner_etal16}, 
the parameters of which are listed in Table~\ref{tab:fidp}.

\begin{table*}
	\centering
	\begin{tabular}{|c|c|c|c|c|c|c|c|}
	\multicolumn{8}{|c|}{\textbf{Fiducial HOD Parameters}} \\
		\hline\hline
	 & $\log M_{\textrm{min}}$ & $\sigma_{\log M}$ & $\alpha$ & $\log M_{\textrm{0}}$ & $\log M_{\textrm{1}}$ & $A_{\textrm{cen}}$ & $A_{\textrm{sat}}$\\
		\hline
		 $M_\textit{r}$ $<$ -19.0 & 11.64 & 0.5119 & 1.040 & 10.25 & 12.80 & 0 & 0 \\
		 $M_\textit{r}$ $<$ -19.5 & 11.75 & 0.4458 & 1.116 & 11.29 & 13.06 & 0 & 0 \\
		 $M_\textit{r}$ $<$ -20.0 & 11.97 & 0.3485 & 1.144 & 11.31 & 13.29 & 0 & 0 \\
         $M_\textit{r}$ $<$ -20.5 & 12.25 & 0.1854 & 1.197 & 11.20 & 13.59 & 0 & 0 \\
         $M_\textit{r}$ $<$ -21.0 & 12.82 & 0.5595 & 1.337 & 11.96 & 13.99 & 0 & 0 \\
		\hline\hline
	\end{tabular}
\caption{In this table, we list the fiducial HOD parameters adopted for each luminosity threshold of galaxies, taken from the fits of \citet{zentner_etal16}. Of the 5 standard HOD parameters, $\Mmin$ is the mass at which a halo has a 
50\% probability of hosting a central galaxy; 
$\sigma_{\log M}$ determines the rate that 
$\langle \Ncen \vert \Mvir \rangle$ transitions from zero 
to unity; $\Mzero$ is the truncating mass, below which no 
satellite galaxies are allowed; 
$\Mone$ is the halo mass at which the mean satellite 
number is unity; and $\alpha$ is the index of the satellite occupation power law. Besides the standard HOD parameters, we also allow $\Acen$ and $\Asat$ to vary, which control the amount of galaxy assembly bias for central and satellite galaxies respectively. In doing this we treat galaxy assembly bias as a deviation from the standard HOD model to be constrained.}
\label{tab:fidp}\end{table*}

Each of the observables is binned in a particular manner. We have selected the 
binning scheme to ensure that our binning does not significantly degrade 
the constraining power of any individual observable. We do this by performing 
a series of analyses in which the bin sizes are reduced in each analysis. 
We choose bin sizes for each observable such that further refinement of the 
bins would not yield significant improvement in parameter constraints.
We specify the range of the independent variable for each observable
observable (for example, in the case of $\wprp$, we take 
$0.1 \le r_{\mathrm{p}}/\Mpch \le 31.6$) and increase the 
number of bins until parameter constraints saturate. 
This process has been described in detail in \citet{hearin_etal12}. 
We find that the constraining power of all observables saturates 
at fewer than 30 bins, so we take 30 bins for all observables 
for simplicity. The binning scheme for which our main results are 
obtained is shown in Table~\ref{tab:binning}.
\footnote{\new{Our results are insensitive to the largest length 
scales included in our analysis because statistics on these scales 
are measured with relatively low signal-to-noise. We have verified 
that excluding the few largest bins of $\wprp$ and $\Dsigma$ from our 
analyses results in negligible quantitative change in our constraints 
(typically below $1\%$, and as large as $\sim 2\%$ in the most extreme cases), 
and no qualitative change to our conclusions.}}

\begin{table*}
	\centering
	\begin{tabular}{|c|c|c|c|c|c|c|}
	\multicolumn{7}{|c|}{\textbf{Observable Bin Definition}} \\
		\hline\hline
	Observable & $\wprp$ & $\Dsigma$ & $\VPF$ & $\Pncic$ & $\Pncia$ & $\Pratio$\\
     Bin No. & $\rp [\Mpch]$ & $\rp [\Mpch]$ & $r [\Mpch]$ & $\Ncic$ & $\Ncia$ & $N_2/N_5$ \\
    \hline
1  & * & 0.11 & 1.00 & $\{0 \}$ & $\{0 \}$ & $\interval[{0.000, 0.033})$ \\
2  & 0.11 & 0.13 & 1.08 & $\{1\}$ & $\{1 \}$ & $\interval[{0.033, 0.067})$ \\
3  & 0.14 & 0.16 & 1.17 & $\{2\}$ & $\{2 \}$ & $\interval[{0.067, 0.100})$ \\
4  & 0.17 & 0.20 & 1.27 & $\{3\}$ & $\{3 \}$ & $\interval[{0.100, 0.133})$ \\
5  & 0.20 & 0.24 & 1.37 & $\{4\}$ & $\{4 \}$ & $\interval[{0.133, 0.167})$ \\
6  & 0.25 & 0.29 & 1.49 & $\{5\}$ & $\{5 \}$ & $\interval[{0.167, 0.200})$ \\
7  & 0.30 & 0.35 & 1.61 & $\{6\}$ & $\{6 \}$ & $\interval[{0.200, 0.233})$ \\
8  & 0.37 & 0.43 & 1.74 & $\{7\}$ & $\{7 \}$ & $\interval[{0.233, 0.267})$ \\
9  & 0.45 & 0.52 & 1.89 & $\{8\}$ & $\{8 \}$ & $\interval[{0.267, 0.300})$ \\
10 & 0.54 & 0.62 & 2.04 & $\{9\}$ & $\{9 \}$ & $\interval[{0.300, 0.333})$ \\
11 & 0.66 & 0.76 & 2.21 & $\interval[{10, 12})$ & $\interval[{10, 12})$ & $\interval[{0.333, 0.367})$ \\
12 & 0.81 & 0.92 & 2.40 & $\interval[{12, 13})$ & $\interval[{12, 14})$ & $\interval[{0.367, 0.400})$ \\
13 & 0.99 & 1.11 & 2.59 & $\interval[{13, 15})$ & $\interval[{14, 16})$ & $\interval[{0.400, 0.433})$ \\
14 & 1.20 & 1.35 & 2.81 & $\interval[{15, 18})$ & $\interval[{16, 19})$ & $\interval[{0.433, 0.467})$ \\
15 & 1.47 & 1.63 & 3.04 & $\interval[{18, 20})$ & $\interval[{19, 22})$ & $\interval[{0.467, 0.500})$ \\
16 & 1.79 & 1.98 & 3.29 & $\interval[{20, 24})$ & $\interval[{22, 26})$ & $\interval[{0.500, 0.533})$ \\
17 & 2.18 & 2.39 & 3.56 & $\interval[{24, 27})$ & $\interval[{26, 30})$ & $\interval[{0.533, 0.567})$ \\
18 & 2.66 & 2.90 & 3.86 & $\interval[{27, 31})$ & $\interval[{30, 35})$ & $\interval[{0.567, 0.600})$ \\
19 & 3.24 & 3.51 & 4.18 & $\interval[{31, 36})$ & $\interval[{35, 41})$ & $\interval[{0.600, 0.633})$ \\
20 & 3.95 & 4.26 & 4.52 & $\interval[{36, 42})$ & $\interval[{41, 48})$ & $\interval[{0.633, 0.667})$ \\
21 & 4.82 & 5.16 & 4.89 & $\interval[{42, 48})$ & $\interval[{48, 57})$ & $\interval[{0.667, 0.700})$ \\
22 & 5.88 & 6.25 & 5.30 & $\interval[{48, 55})$ & $\interval[{57, 66})$ & $\interval[{0.700, 0.733})$ \\
23 & 7.17 & 7.57 & 5.74 & $\interval[{55, 64})$ & $\interval[{66, 78})$ & $\interval[{0.733, 0.767})$ \\
24 & 8.75 & 9.17 & 6.21 & $\interval[{64, 74})$ & $\interval[{78, 91})$ & $\interval[{0.767, 0.800})$ \\
25 & 10.6 & 11.11 & 6.72 & $\interval[{74 ,85})$ & $\interval[{91, 106})$ & $\interval[{0.800, 0.833})$ \\
26 & 13.0 & 13.46 & 7.28 & $\interval[{85 ,98})$ & $\interval[{106, 125})$ & $\interval[{0.833, 0.867})$ \\
27 & 15.8 & 16.30 & 7.89 & $\interval[{98 ,113})$ & $\interval[{125, 146})$ & $\interval[{0.867, 0.900})$ \\
28 & 19.3 & 19.75 & 8.53 & $\interval[{113, 130})$ & $\interval[{146, 171})$ & $\interval[{0.900, 0.933})$ \\
29 & 23.6 & 23.93 & 9.24 & $\interval[{130, 150})$ & $\interval[{171, 200})$ & $\interval[{0.933, 0.967})$ \\
30 & 28.7 & 28.99 & 10.00 & $\interval[{150,
+\infty})$ & $\interval[{200,+\infty})$ & $\interval[{0.967, 1.000}]$ \\
		\hline\hline
	\end{tabular}
\caption{Definition of bins for each observable. We show the values of bin centres for the bins in which $\wprp$ and $\Dsigma$ are measured, the set of radii of spheres used for evaluating $\VPF$, the intervals defining each bin in the histograms of counts-in-cylinders and annuli, and the ratio $N_2/N_5$. We use the same number of bins for each observable, such that our comparison of constraining power is not sensitive to bin number. * Note that the number density $\ngal$ is listed as the first bin of $\wprp$ in this table, but it is included in the analysis for all possible combinations of observables.}
\label{tab:binning}\end{table*}

\subsubsection{Projected Two-Point Correlation Function, $\wprp$}
\label{sec:wp}

The projected two-point correlation function, $\wprp$, is a 
canonical observable that has been considered in 
numerous previous analyses to inform halo occupation \citep[e.g.,][]{vdBosch07, zheng09, zehavi_etal11}. 
It is defined by 
\begin{equation}
\label{eq:wp}
\wprp = 2\int_{0}^{\pimax}d\pi \ \xi(\rp,\pi)
\end{equation}
where $\xi(\rp,\pi)$ is the excess probability of finding galaxy pairs with projected and line-of-sight separations $\rp$ and $\pi$, respectively.
We estimate $\wprp$ from our mock catalogues
by counting galaxy pairs that have a 
projected separation in a bin 
of $\rp$ within a perpendicular 
distance of $\pimax$ in redshift space.
We choose $\pimax = 60\Mpch$, as is done by \cite{zehavi_etal11}, 
according to whom this integration limit is large enough to include 
most correlated pairs and minimize the impact of the details of peculiar 
velocity models, yet sufficiently small to give a stable result by 
suppressing noise from very distant, uncorrelated pairs. 
We compute $\wprp$ in 29 logarithmically spaced radial bins from 
$\rp = 0.1\, \Mpch$ to $\rp =31.6\, \Mpch$. 

The projected two-point clustering 
of our fiducial models are shown for two luminosity threshold samples
in the upper, left panel of Fig.~\ref{fig:obs_example}. 
The figure exhibits several well-known characteristics of 
galaxy clustering. First, brighter galaxies cluster more 
strongly. Second, the galaxy two-point correlation function can roughly 
be described as a power law, $\wp = (\rp/r_0)^{\alpha}$, 
with index $\alpha \approx -0.8$. Third, in more detail, 
the correlation function exhibits a small deviation from a
power law near $\rp \sim 2\, \Mpch$
which is due to the transition from 
galaxy pairs that reside in distinct haloes (the ``two-halo'' 
 term) on large scales ($\rp \gtrsim 2\Mpch$) and 
pairs of galaxies that reside in a common halo 
(the ``one-halo term'') on scales $\rp \lesssim 1\Mpch$.

\subsubsection{Galaxy--Galaxy Weak Lensing, $\Dsigma$}
\label{sec:ds}

In addition to the projected two-point clustering, galaxy--galaxy 
weak lensing is another observable statistic that has been used by 
many previous authors to constrain halo occupation from observational 
data \citep[e.g.,][]{mandelbaum06b, cacciato_etal09, vUitert.etal.16}. The canonical observable, $\Dsigma$, 
is the excess surface density of mass around galaxies 
projected along the line-of-sight, and averaged over all potential 
lens galaxies in the sample,
\begin{equation}
\label{eq:ds}
\Dsigma = \bar{\Sigma}(<\rp) - \Sigma(\rp)
\end{equation}
where $\Sigma(\textit{r}_\textrm{p})$ is the projected surface density evaluated at position $\rp$ relative to the centre of the lens galaxy, 
and $\bar{\Sigma}(<\rp)$ is the mean projected, two-dimensional, surface mass density 
within a projected distance or $r_{\mathrm{p}}$ from the 
lens galaxy.
We compute $\Dsigma$ in 30 logarithmically-spaced 
radial bins from 
$\rp = 0.1 \, \Mpch$ to $\rp = 31.6 \, \Mpch$. 
The simulations that we use are gravity-only N-body simulations, 
so our estimates of $\Dsigma$ include neither baryonic mass nor 
any influences of baryons on the dark matter distribution 
\citep[e.g.,][]{rudd_etal08}.

The galaxy--galaxy lensing signal in our fiducial catalogues 
is depicted in the upper, right panel of Fig.~\ref{fig:obs_example}. 
It is evident from this panel that galaxies in the higher-luminosity 
samples are more strongly correlated with mass, indicating the 
well-known fact that more luminous galaxies tend to reside in 
more massive dark matter haloes. The feature due to the transition 
between the one-halo and two-halo terms in the galaxy matter 
correlation function is evident near $\rp \sim 2\, \Mpch$ as well.

\subsubsection{Void Probability Function $\VPF$}
\label{sec:vpf}

We examine a number of options for observables in addition to the 
canonical $\wprp$ and $\Dsigma$, among them the VPF. The VPF has 
been examined in previous studies on assembly bias with mixed 
conclusions \citep{tinker_etal13,zentner_etal14} and, 
in principle, depends upon all of the n-point functions
\citep[e.g.,][]{peebles_book80}. 
To estimate $\VPF$, we randomly place spheres of radius 
$r$ throughout our simulation volume 
and enumerate the probability of the spheres 
containing zero galaxies (and thus being classified as voids) as  
\begin{equation}
\label{eq:vpf}
\VPF = \frac{N_{\textrm{void}}(r)}{N_{\textrm{sphere}}(r)},
\end{equation}
where $N_{\mathrm{sphere}}$ is the total number of spheres that we use for 
the estimate ($N_{\mathrm{sphere}}=10^5$ in this work) and 
$N_{\textrm{void}}$ is the number of spheres that are found 
to enclose zero galaxies. We compute $\VPF$ at 30 
logarithmically-spaced radii from 
$r = 1 \, \Mpch$ to $r = 10 \, \Mpch$. 
We remind the reader that these calculations 
are performed in redshift space, by mapping the coordinates of galaxies according to their line-of-sight velocities.

The $\VPF$ of our fiducial models are depicted in the left, middle panel of 
Fig.~\ref{fig:obs_example}. As with $\wprp$ and $\Dsigma$, several expected 
features of the $\VPF$ are evident. The VPF drops from nearly unity 
on small scales to well below unity beyond a scale of $r \sim 10\, \Mpch$ 
and voids are more likely for higher luminosity galaxy samples, due 
largely to their overall lower number density.

\subsubsection{Counts-in-cylinders (CIC) Statistic $\Pncic$}
\label{sec:cic}

Galaxy counts, particularly counts of galaxies within cylindrical 
volumes in redshift space, have been studied for decades 
\citep[e.g.,][]{fry1978,Balian1988,jean-michel_etal90,Baugh1995,colombi_etal95,Szapudi1996,kim_strauss98,Hogg2004,Kauffmann2004,Blanton2006,Barton2007,reid_spergel09,Berrier2011,oguri_lin15,gruen_des_lensing_cic2018}. 
The average number of companions that a 
galaxy will have within a particular cylinder 
can be computed from the two-point correlation function; 
however, the distribution of counts-in-cylinders depends, at least in principle, 
upon all of the higher n-point functions \citep{peebles_book80} and can complement the 
two-point function as a study of the galaxy halo relationship.

We compute counts in cylinders (CIC) from our galaxy catalogues as 
follows. We centre a cylinder of transverse radius $\rcic$ and 
depth $\pm \Delta v$ (in redshift space)
on each galaxy in the sample and count 
the number of companion galaxies that fall within the 
cylinder. This procedure enables us to estimate a probability distribution 
of companion number, $\Pncic$, which is the probability that any 
galaxy has $\Ncic$ companions within the cylinder. For the primary 
results that we present in this paper, we use $\rcic = 2 \Mpch$ and a maximum relative velocity of 
$\Delta v = 1000\, \kms$, corresponding to a half-length of
$L = 10 \Mpch$, assuming velocities are only due to the
Hubble flow. We choose cylinders of a 
transverse radius $\rcic$ on the order of a few $\Mpch$ 
in order to include galaxy companions separated by a scale 
on which assembly bias is known to introduce a distinct feature 
in halo clustering \citep{hearin_etal16,zentner_etal16,sunayama_etal16}.  
We have experimented with a variety of alternative
cylinder radii and depths, 
finding that our results remain qualitatively similar. 
When characterizing the PDF $\Pncic$,
we count the first few values of $\Ncic$ 
individually and group larger $\Ncic$ values into 
logarithmically spaced bins, as indicated in Table~\ref{tab:binning}.

Examples of our counts-in-cylinders distributions, $\Pncic$, 
for two of our luminosity threshold samples, can 
be seen in the middle, right-hand panel of Fig.~\ref{fig:obs_example}. As is expected, 
the probability of having a large number of companions in a cylindrical 
cell increases dramatically with decreasing galaxy luminosity due to the 
higher number density of galaxies with lower luminosities.

\subsubsection{Counts-in-annuli (CIA) Statistic $\Pncia$}
\label{sec:cia}

To complement counts-in-cylinders, we also examine counts of neighbour 
galaxies in annuli. The counts-in-annuli (CIA) enable one to get a sense 
of clustering as a function of scale and to compare smaller-scale, 
intra-halo clustering to larger-scale clustering. The statistic $\Pncia$ 
is the probability that the number of companions within the annulus 
is equal to $\Ncia$, analogous to $\Pncic$. Unlike the counts-in-cylinders statistic, which roughly probes the halo-occupation statistics on the ``one-halo'' scale, $\Pncia$ is a novel statistic, introduced here, and specifically designed to probe the immediate, supra-halo environments of galaxies.

As with CIC, we choose fixed dimensions for the annuli that we use and 
explore the constraining power of the distribution of counts around 
galaxies in our catalogues. Our annuli have inner radii of 
$\rinner = 2 \, \Mpch$ and outer radii of $\router =  5\, \Mpch$. 
As with CIC, the annuli have a depth in the redshift dimension of 
$10\, \Mpch$, 
corresponding to a velocity difference of 
$\Delta v = 1000 \, \kms$. As with CIC, this geometry is chosen in order to 
probe the immediate environments of haloes, particularly on scales 
where assembly bias has already been shown to induce a feature in 
galaxy clustering \citep[e.g.,][]{hearin_etal16,sunayama_etal16}. 
We have experimented with moderately different annular dimensions 
and obtained qualitatively similar results in all cases. 
We group values of $\Ncia$ in a similar way to $\Ncic$, 
as detailed in Table~\ref{tab:binning}. 

Examples of $P(\Ncia)$ for our fiducial catalogues are given in 
the lower, left-hand panel of Fig.~\ref{fig:obs_example}. The 
CIA distribution shares most of the qualitative features of 
the CIC distribution, though the counts are generally 
higher because the volumes of our annuli exceed the volumes 
of our cylinders by a factor of $\sim 5$. 

\subsubsection{Distribution of Cylinder Count Ratios, $P(N_2/N_5)$}

As a distinct way of characterising the clustering environments of 
galaxies, we also consider the distribution of the ratio of two cylinder 
counts. The first count is within a 
cylinder with a radius of $\rcic = 2\, \Mpch$ and the second, 
larger cylinder has a radius $\rcic = 5\, \Mpch$ for each galaxy.
For both cylinders we adopt the same depth, $\Delta v = 1000 \, \kms$
as for the CIC and CIA statistics discussed above. For each galaxy in our 
catalogues, we compute the companion counts $N_2$ and $N_5$ within each of 
these cylinders, and take the probability distribution of the ratio 
of these two numbers as the 
statistic of interest (notice that the inner cylinder is the same 
cylinder used in our CIC calculations, so that $N_5 = N_2 + \Ncia$. 
Similar to $\Ncia$, this is a novel statistic to probe the large scale distribution of galaxies that, to the best of our knowledge, has not been utilized before.
The intention of this statistic is to probe 
the relative clustering within a halo (the ``one-halo term'') to that in its immediate vicinity. 
We measure the probability distribution of this ratio, $\Pratio$ in 30 
linearly-spaced bins from 0 to 1. 

Probability distributions of the cylinder count ratio are shown in the lower, 
right-hand panel of Fig.~\ref{fig:obs_example}. 
It is evident that the higher luminosity sample 
has a distribution with more discreteness noise, 
as a result of the low number density. 
The more luminous sample has a significantly 
higher probability of having low values of 
$N_2/N_5$ than the lower luminosity sample. 
This is due to the fact that satellite galaxies 
are increasingly rare in the higher-luminosity 
samples, so that $N_2$ is increasingly likely to 
be either small or zero in such samples compared 
to lower-luminosity samples (see the 
{\rm right, middle} panel of Fig.~\ref{fig:obs_example}).

\subsection{Fisher Analysis}
\label{sec:fisher}

We use a Fisher matrix analysis 
\citep{bond_jaffe_knox98,dodelson_book03} 
to forecast the constraining
power of each of the observables 
described above and combinations thereof.
Despite its approximate nature, we elect to use a
Fisher matrix due to the computational expense of utilizing 
other techniques. Employing a technique that directly quantifies 
uncertainty by averaging the posteriors found from each mock catalogue, 
either via Markov Chain Monte Carlo \citep{goodman_weare10,emcee}
or Approximate Bayesian Computation \citep{weyant_etal13}, 
to perform the very large number of analyses
that we undertake using mock catalogues is substantially 
more computationally intensive.

The Fisher matrix is a measure of the ideal amount of 
information that can be obtained from an experiment. 
The Fisher matrix is defined as
\begin{equation}
\label{eq:fisher}
\mathbf{I} = \frac{\partial \mathbf{f}^\textrm{T}}{\partial \mathbf{p}}\mathbf{C}^{-1}\frac{\partial \mathbf{f}}{\partial \mathbf{p}}  , 
\end{equation}
where $\mathbf{f}$ is the set of observables, 
$\mathbf{C}$ is the covariance of the observables, 
and $\mathbf{p}$ is the parameter set to be constrained. 
The set of observables $\mathbf{f}$ includes all 
bins of each observable quantity (e.g., 29 bins of 
$w_{\mathrm{p}}(r_{\mathrm{p}})$, 
30 bins of $\Delta \Sigma(r_{\mathrm{p}})$, etc.), 
while $\mathbf{p}$ represents the set of all 
model parameters. 
The notation $\partial \mathbf{f}/\partial \mathbf{p}$ 
represents the matrix of values constructed by differentiating each observable 
with respect to each of the individual parameters, so that 
the matrix element $\partial f_{\mathrm{i}}/\partial p_{\mathrm{j}}$ 
represents the derivative of the $i^{\mathrm{th}}$ observable 
with respect to the $j^{\mathrm{th}}$ model parameter. 
Both the derivatives and the covariance are evaluated at a 
single, {\em fiducial} point in the parameter space, which 
is assumed to be the true underlying model.

The expected $1\sigma$ error on any inferred parameter, 
marginalized over all other parameters,
can be obtained by taking the square root of the corresponding 
diagonal term of the posterior covariance matrix $\bm{\Sigma}$, 
which is the inverse of the Fisher matrix. 
Hence, the forecasted marginalized uncertainties 
in the parameters are
\begin{equation}
\label{eq:sigma}
\bm{\sigma} = \sqrt[]{\textrm{Diag}(\bm{\Sigma})} = \sqrt[]{\textrm{Diag}(\mathbf{I}^{-1})}.
\end{equation}
In our study of the constraining power of various observables, 
we explore the 6 dimensional parameter space, 
spanned by 
$\alpha$, $\log{M_\mathrm{1}}$, $\sigma_{\log{M}}$,  $\log{M_{\rm{min}}}$, $A_{\rm{cen}}$, 
and $A_{\rm{sat}}$. The parameter $\log{M_\mathrm{0}}$ 
is part of both the standard HOD and the dHOD models. 
However, we set $\log{\Mzero}$ to its fiducial value in our 
analyses and do not allow it to vary. We do this because 
$\log{\Mzero}$ is poorly constrained by 
these data \citep[see][]{zentner_etal16}.

The fiducial points about 
which we evaluate our Fisher matrices are given by
the parameters listed in Table~\ref{tab:fidp}, 
and differ for each luminosity threshold sample.
In the next two subsections, 
we discuss the computation of the covariance matrix, 
$\mathbf{C}$, and the derivatives of the observables, 
$\partial \mathbf{f}/\partial \mathbf{p}$, respectively.

\subsection{Covariance}
\label{sec:cov}

\begin{figure*}
\centering
  \includegraphics[width=\textwidth]{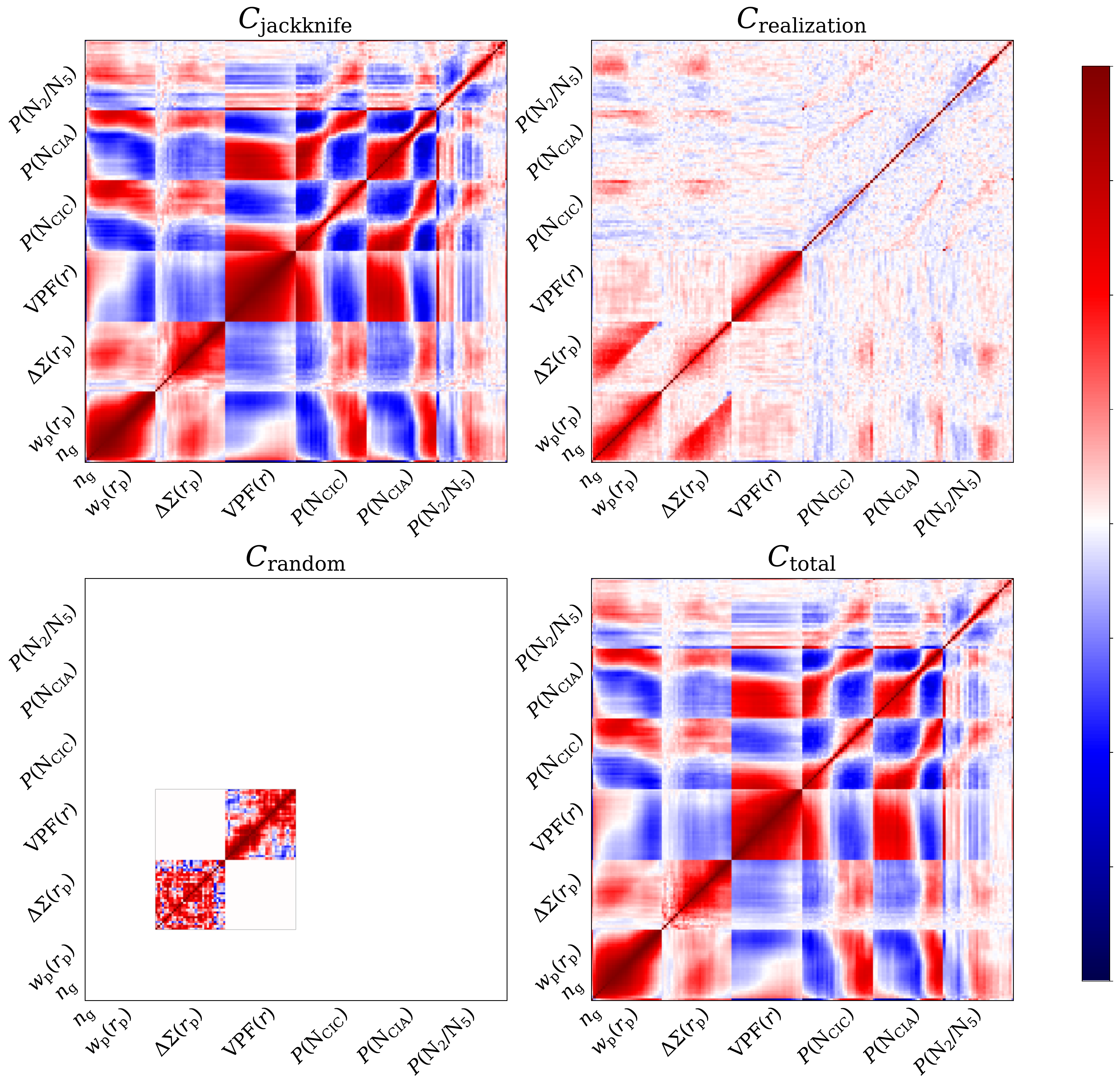}
  \caption{\label{fig:corr}In this figure, we show the total covariance matrix 
  along with the three matrices that we sum in order to compute the total 
  covariance matrix, each normalized to correlation matrices, for the $M_r < -19.0$ threshold sample. $\Creal$ and 
  $\Crandom$ are computed using 
  our fiducial HOD parameters. The sample variance contribution, 
  $\Cjack$, is computed using a mock catalogue based on abundance matching 
  in order to eliminate the stochasticity associated with any individual 
  mock HOD catalogue. $\Cjack$ is the major contribution, while $\Creal$ and $\Crandom$ are subdominant. Covariances for other thresholds are 
 qualitatively similar. Blocks of these matrices corresponding to 
 the bins of a specific type of observable (e.g., $\wprp$ or $\Pncic$) 
 are labelled as such. Each such block contains 30 rows and columns 
 corresponding to the 30 bins used for each observable.}
\end{figure*}

In order to implement the Fisher approximation for the marginalized 
constraints on model parameters, we must compute a covariance matrix 
about the fiducial point in the parameter space. The covariance matrix 
that we calculate has three contributions. The first, and dominant, 
contribution is from sample variance (sometimes called ``cosmic'' 
variance in this context). We estimate the sample variance contribution 
using jackknife resampling of the simulation volume, while recognizing the caveat 
that jackknife resampling is known to underestimate covariances. 
We will refer to this component of the covariance as $\Cjack$. 
The second contribution 
to the covariance matrix is due to the stochasticity of populating a 
simulation with galaxies drawn from the probability distribution functions of 
the (d)HOD. Multiple realizations of the same underlying model in 
identical, finite volumes will lead to mildly different predictions 
due to this stochasticity. 
We refer to this contribution to the covariance as $\Creal$. 
Third, we use a fixed set of randomly distributed centres of 
spheres in the calculation of $\VPF$ as well as a fixed subsample of dark 
matter particles in the calculation of $\Dsigma$. These choices 
introduce a small contribution to the covariance that we denote 
$\Crandom$. The total covariance matrix that we use 
is the sum of each of these contributions
\begin{equation}
\Ctot = \Cjack+\Creal+\Crandom
\label{eq:covtot}.
\end{equation}
As an example, Fig.~\ref{fig:corr} depicts the covariance matrices $\Cjack$, $\Crandom$, $\Creal$ and their linear combination $\Ctot$, for the $M_r < -19.0$ threshold sample, as normalized correlation matrices.

The contributions from $\Creal$ and 
$\Crandom$ are straightforward to compute. To estimate 
$\Creal$, we populate the halo catalogue with the fiducial HOD multiple times, 
each time using a new random seed, and compute the covariance across the 
measurements from the resultant mock galaxy catalogues.
$\Creal$ is displayed in the upper right panel of Fig.~\ref{fig:corr}. For 
$\Crandom$, we repeat measurements of $\Dsigma$ and $\VPF$ on the same mock galaxy catalogue, but with different sets of particle subsamples and sphere centres, respectively, and calculate the covariance. 
Note that this matrix only consists of the two corresponding blocks, 
as shown in the lower left panel of Fig.~\ref{fig:corr}.

Finally, to estimate $\Cjack$, which is shown in the upper left panel of 
Fig.~\ref{fig:corr}, we divide the simulation box 
into 10$\times$10 cuboid cells, each of size 
$25\, \Mpch \times 25\, \Mpch \times 250\, \Mpch$. 
The long axes of each cuboid are the same as the length of
the simulation volume and are assumed to lie along the line of sight. 
For each mock catalogue, we construct three such sets of jackknife samples by 
choosing, in turn, the $x$, $y$, and $z$ dimensions of the simulation 
cube as the line-of-sight direction. Our final covariances are the 
averages of the three covariances computed for each of the three 
projections. We construct this average to minimize the contributions 
from any significant variations that may, by chance, 
fall along any individual projection. For each set of jackknife 
samples coming from each of the three projections of the mock 
catalogues, we exclude individual jackknife cells in turn, and compute 
the jackknife contribution in the usual manner \citep{jackknife-jones1956,jackknife-quenouille1956,jackknife-tukey1958,astrostat_wall_jenkins_2003}.

For the purposes of computing jackknife covariances only, the mock 
catalogues that we use are {\em not} based on our fiducial HODs.  
Our jackknife covariance mock catalogues are based upon abundance matching with zero 
scatter \citep{kravtsov04a}. We construct these catalogues by populating 
haloes that have 
the highest values of $V_{\mathrm{peak}}$, 
with galaxy number densities consistent with 
HOD realizations. This modification is necessary for the following reason. 
Each mock HOD-based catalogue is a realization of the underlying HODs. 
Therefore, there is inherent stochasticity in the covariance matrix estimates. 
Moreover, in the HOD formalism, each luminosity threshold must be treated 
independently, which, in turn, means that the covariances in different threshold 
samples can fluctuate independently. This makes comparing covariances across luminosity 
thresholds challenging because to do this using the HOD approach requires marginalizing 
the stochasticity over a very large number of mock catalogues. 
The abundance matching approach that we have adopted allows us to 
circumvent this difficulty because there is no stochasticity in the 
mock catalogues. 
Therefore, the stochasticity associated with building mock catalogues does 
not contribute to our $\Cjack$ estimates. 
This ensures that our $\Cjack$ estimates 
vary smoothly with the luminosity threshold of the sample. We have found that this procedure reduces the noisiness of our forecasts, 
yet does not alter our qualitative results.

All six of our candidate observables are based on pair or neighbour 
counting, which reduces the choice 
of algorithm to determining which counts 
to exclude for each jackknife subsample. 
For $\wprp$, we discard a pair if either or both of the galaxies 
reside in the excluded cell. For $\Dsigma$, we only calculate the dark 
matter density profile around galaxies that live outside the excluded cell. 
Note, though, that in doing so we include dark matter 
particles that lie in that cell. Excluding such particles 
would lead to anomalous density profiles that are not easily corrected 
because the subsampling procedure violates the periodicity of the 
simulation volume. 
Similarly, for $\VPF$, we place random spheres about points outside 
of the excluded cell; however, for the purposes of determining whether or 
not a particular sphere is a void region, galaxies within the excluded 
cell are taken into account. 
And for the count statistics, $\Pncic$, $\Pncia$, 
and $\Pratio$, we only centre cylinders on galaxies outside the 
excluded cell, but include companion galaxies within the excluded cell in our 
counts.

Of the three contributions, $\Cjack$ is the dominant component, 
and $\Crandom$ is negligibly small, suggesting that we have used 
sufficiently large samples of VPF centres and dark matter particle positions 
to render the noisiness induced by finite sampling of these distributions 
negligibly small. 

Direct inversion of the covariance matrix $\mathbf{C}$ is problematic 
numerically. Briefly stated, the uncertainty in the covariance will lead to
the smallest eigenvalues of $\mathbf{C}$ being dominated by noise. When 
a matrix is inverted, its eigenvalues are inverted, 
which implies that the small, noisy
eigenvalues of $\mathbf{C}$ become the large and 
noisy eigenvalues of its inverse. The inverse hence becomes dominated 
by this noise.
This problem is further compounded by the large differences among the matrix elements inherited from the differences between the 
natural scales of the different observables, leading to extremely 
large differences in the sizes of the matrix eigenvalues. For example, 
the natural scale of the two-point function, $\wprp$, is $\sim 10^2$, whereas 
the natural scale of $\Pncic$ is on the order of $\sim 10^{-2}$ (see Fig.~\ref{fig:obs_example})
and this difference leads to very different covariance matrix elements.

A common approach when faced with this problem is to truncate the 
smallest eigenvalues of $\mathbf{C}$, and calculate 
the Moore-Penrose pseudo-inversion \citep{pinv_penrose_1955}. 
We therefore normalize $\mathbf{C}$, 
and perform pseudo-inversion, excluding the eigenvalues of the correlation matrix
smaller than $10^{-5}$ times the largest eigenvalue. Choosing other 
reasonable values of this cutoff does not substantially impact our results.
In the future, when larger data sets are available, and the noise level of the 
covariance is sufficiently low, pseudo-inversion may no longer be necessary.

Our covariance matrices have several noteworthy features. 
Firstly, from an inspection of Fig.~\ref{fig:corr}, 
we see strong self correlation between the bins of $\wprp$, $\Dsigma$, 
and $\VPF$ over a wide range of scales. 
On the contrary, $\Pratio$ shows weak correlation among its bins 
and with other observables, as it measures the distribution 
of the dimensionless ratio $N_2/N_5$, and is insensitive 
to the cosmic variance of galaxy number density. 
Secondly, for $\Pncic$ and $\Pncia$, the probability of smaller 
counts and larger counts sum up to unity, and are anti-correlated by construction, 
producing the sign reversal in the corresponding matrix blocks.
Additionally, when comparing $\Cjack$ and $\Creal$, 
it is obvious that the observable values are more correlated 
among jackknife subsamples than stochastic realizations.
The sign of correlation coefficients approximately 
coincide between the two contributions, 
with the exception of blocks involving $\VPF$. 
In $\Creal$, $\VPF$ has a weak positive correlation with 
$\wprp$ and $\Dsigma$ in most of the bins, 
while in $\Cjack$, $\VPF$ is anti-correlated with 
$\wprp$ at larger $\rp$ and $\Dsigma$. 
These are non-trivial effects, as $\VPF$ is dependent 
on multiple moments of the galaxy number density field. 
The jackknife subsamples probe different regions of the box, 
with denser regions corresponding to stronger 
galaxy--galaxy and galaxy--matter correlation as well as 
fewer voids, leading to the anti-correlation in $\Cjack$.
On the other hand, among different realizations, 
higher values of $\wprp$ and $\Dsigma$ result not 
from higher galaxy number densities but when more 
galaxies are concentrated in clusters, allowing more 
voids to exist in the rest of the space, giving rise 
to a positive correlation with $\VPF$.
Covariances for other luminosity samples have qualitatively similar features. 

\new{When applying the same analyses to galaxy survey data, observational uncertainties need to be taken into account, here we discuss how our covariances compare to observational covariances from SDSS data.
For statistics that depend only on galaxy distribution, e.g., all of our observables except $\Dsigma$, the observational uncertainties depend on the survey volume and target number density to first order, both of which we mimic in our analyses.
In comparison with the covariances from SDSS, which has a similar volume for the $M_r<-20.0$ sample to the Bolshoi Planck simulation, indeed our jackknife covariance matrix for $\wprp$ is comparable to the measurement in \citet{zehavi_etal11}.
For the fainter samples, SDSS has smaller volumes and hence larger covariances than ours, and vice versa for the brighter samples. 
Similarly, we expect the other observables to have covariances comparable to SDSS observation, with the exception of $\Dsigma$.
Aside from the first-order comparability, there are other factors that need to be accounted for in observation.
Fibre collision \citep[see, e.g.,][]{o'mill2012} affects pairs with small separations, i.e., the smallest bins of $\wprp$ and $\Dsigma$, as well as $\Pncic$ and therefore $\Pratio$. This effect needs to be forward modelled in survey data analyses.
However, $\Pncia$ does not depend on the closest galaxy companions, and is robust to fibre collision. As we will show in Section~\ref{sec:results}, $\Pncia$ typically provides tight constraints comparable to $\Pncic$.
Some of the other second-order observational uncertainties are blending and saturation in target selection, redshift measurement errors, and geometric features, all of which we expect to be subdominant in this context.}

\new{On the other hand, the covariances for $\Dsigma$ measured for SDSS data \citep{mandelbaum06a} are significantly larger than our covariances. 
This is expected because the survey data is dominated by shape noise, 
while we neglect shape noise in our study, assuming infinite source densities. 
Taking non-zero shape noise into account will result in weaker 
constraints from lensing than those that we find here, 
and our forecasts therefore must be regarded as 
the upper limit of constraining power that can be achieved with $\Dsigma$.}

\subsection{Derivative Fitting}
\label{sec:deriv}

In order to compute the elements of the Fisher matrix, it is necessary 
to estimate the partial derivatives of the observables with respect to the 
parameters in the neighbourhood of the fiducial point in the parameter space. 
We designated these derivatives as $\partial \mathbf{f}/\partial \mathbf{p}$ 
in Eq.~(\ref{eq:fisher}) above. Assessing these derivatives from 
realizations of the perturbed models is non-trivial due to the 
inherent stochasticity of using mock catalogues based upon the 
direct population of N-body simulations. Therefore, we give a detailed 
description of our approach to estimating derivatives 
in Appendix~\ref{sec:app_deriv}.

\section{Results}
\label{sec:results}

We perform Fisher matrix analyses in order to 
forecast the constraints on the dHOD 
model that can be extracted from combinations of the galaxy observables 
described in Section~\ref{sec:obs}. 
In Table~\ref{tab:fidp}, we list the fiducial HOD parameters corresponding 
to 5 galaxy samples selected by luminosity. In this section, we present 
our primary results in terms of estimated posterior 1$\sigma$ constraints 
for the four lower luminosity samples that we have studied. 
We exclude the brightest, $M_r<-21.0$ sample from our primary 
results because the results from this sample are subject to excessive
statistical fluctuations due to the relatively small number of 
galaxies above this luminosity threshold within the volume of the 
Bolshoi Planck simulation. For completeness, the results 
from the $M_r<-21.0$ sample are included in our comprehensive 
list of results in Appendix~\ref{sec:app_constraint}.

Before proceeding to our results, we note that the 
dominant contribution to our errors are from sample 
variance due to the finite volume of the Bolshoi Planck 
simulation (see Section~\ref{sec:cov} above). The 
absolute constraints on parameters will decrease with 
increasing volume. As such, we focus on {\em relative} 
constraints on parameters from different combinations 
of observables, rather than on the absolute values. 
Our study requires that haloes be resolved with a very large number 
of particles which, in turn, stipulates the use of a high-resolution,
relatively smaller volume simulation such as Bolshoi Planck.

\subsection{Assessing the Complementarity of Observables}

\begin{figure*}
\centering
  \includegraphics[width=\textwidth]{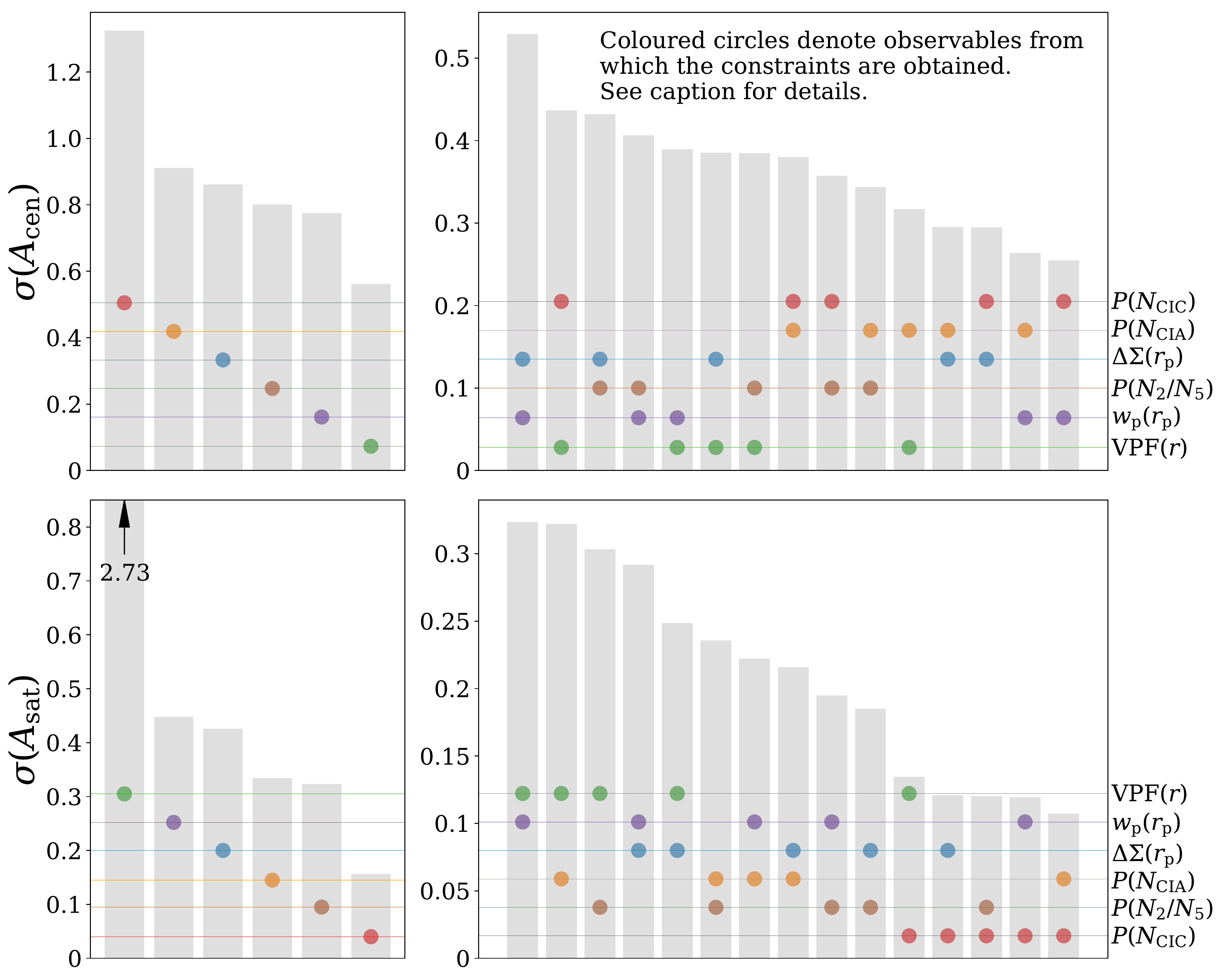}
  \caption{\label{fig:vdB_4p}
  In this figure, we compare the constraints on $\Acen$ and $\Asat$ from the individual observables (left-hand panels) and all the possible combinations of two observables (right-hand panels), using the $M_r<-20.0$ sample as an example. The top row shows the constraints on $\Acen$ and the bottom row shows the constraints on $\Asat$. Note that each of the four panels has a different $y$-axis range. In each panel, we arrange the columns from least constraining (at left) to most constraining (at right). The filled circles in different colours indicate the observables used to compute the constraints of the corresponding column, as labelled on the right. The relative heights of the circles are ordered by the constraining power from each {\em individual} observable on the relevant parameter, shown in the left panels. The absolute heights of each coloured circle do not correspond to the absolute constraints from that individual 
  observable. In the bottom left panel, the black arrow indicates that the individual constraint from $\VPF$ on $\Asat$, the value of which is shown below the arrow, greatly exceeds the range of the $y$-axis.}
\end{figure*}

We begin with a discussion of our forecast constraints from {\em individual} 
observables and combinations of any {\em two} observables. 
We use the $M_r<-20.0$ sample as an example in Fig.~\ref{fig:vdB_4p}, 
to compare the constraints on $\Acen$ and $\Asat$ from the individual 
observables and all the possible combinations of two observables. 
We caution that each of the four panels has a different $y$-axis range, 
which the reader must account for when comparing 
results among different panels.

We display constraints from the individual observables on $\Acen$ 
in the upper left panel of Fig.~\ref{fig:vdB_4p} 
and $\Asat$ in the lower left panel of Fig.~\ref{fig:vdB_4p}. 
The columns are ordered according to increasing constraining 
power, and the filled circles indicate the observables from which 
the constraints were derived. For example, the leftmost column 
in the upper left panel of Fig.~\ref{fig:vdB_4p} shows the 
constraint on $\Acen$ derived from $\Pncic$.

The constraints from individual observables, displayed in the left 
panels of Fig.~\ref{fig:vdB_4p} convey several points. 
The observable $\Pncic$ more 
strongly constrains the assembly bias of satellite galaxies than 
that of central galaxies. 
This is expected because $\Pncic$ primarily probes 
the ``one-halo term'', and is sensitive to the satellite 
population that accompany centrals. For the brighter samples 
with higher satellite fractions, $\Pncia$ is more dependent 
on the satellite population, and constrains $\Asat$ more strongly, 
while its constraining power decreases for $\Acen$. 
The constraints from $\wprp$ are dominated by the smaller radial bins, 
which have higher signal to noise ratios than the measurements at larger scales. 
This causes the absolute constraints on $\Acen$ to be weaker than those on $\Asat$. 
However, with larger volumes, e.g., DESI \citep{levi2013_desi}, 
large-scale clustering will be measured with higher precision, 
enabling better constraints on $\Acen$. We also find that $\VPF$ 
gives strong constraints on $\Acen$, but is extremely inefficient 
in constraining $\Asat$. This can be explained by the fact that 
a single galaxy suffices to eliminate the possibility that a region 
could be a void. The vast majority of satellite galaxies reside 
in haloes where there are central galaxies, which already 
eliminate the void, therefore the void probability function is 
largely insensitive to the abundances of satellite galaxies. 
For this reason, $\VPF$ is a poor probe of not only $\Asat$, 
but all of the parameters that determine satellite populations 
(see Tables~\ref{tab:sigma190}-\ref{tab:sigma205} 
in Appendix~\ref{sec:app_constraint}).

The constraints from the combinations of two observables are 
shown in the right-hand panels of Fig.~\ref{fig:vdB_4p}. 
The columns are again ordered from least constraining 
to most constraining, and the observables used in each 
analysis are marked by filled circles. 
For example, the leftmost column in the upper right panel 
of Fig.~\ref{fig:vdB_4p} shows constraints derived from the 
combination of $\wprp$ and $\Dsigma$. 
The combination of $\wprp$ and $\Dsigma$ has been used in a number of
previous studies; however, we find that this is one of the
least constraining of the combinations that we have
considered for both $\Acen$ and $\Asat$. As we discuss further below, 
this is chiefly because $\wprp$ and $\Dsigma$ share largely common
degeneracies among the (d)HOD parameters, so that combining 
the two does not result in significant decreases in 
marginalized constraints that can be expected when 
combining highly complementary data.

The primary result to be gleaned from Fig.~\ref{fig:vdB_4p} 
is the overall efficacy
of the count-based observables, i.e., $\Pncic$ and $\Pncia$,
to complement either $\wprp$ or $\Dsigma$ 
to constrain the galaxy assembly bias parameters.
We find that when used in combination with count-based statistics, 
$\wprp$ typically outperforms $\Dsigma$. Moreover, it is worth noting that our 
lensing covariance assumes an infinite density of lensing sources, 
so lensing constraints realized from a real survey analysis will be 
further diluted by shape noise contributions to the covariance, 
as we have discussed in Section~\ref{sec:cov}. Therefore, we suspect that this general result will 
be robust to actual survey analyses.
While Fig.~\ref{fig:vdB_4p} displays only
constraints on $\Acen$ and $\Asat$ from a single luminosity threshold sample, we find that 
these qualitative results hold for all thresholds samples considered here (see Fig.~\ref{fig:Asigma} 
and Appendix~\ref{sec:app_constraint}).

\subsection{Complementarity with Clustering}

\begin{figure*}
\centering
\begin{minipage}{0.8\textwidth}
   \includegraphics[width=1\linewidth]{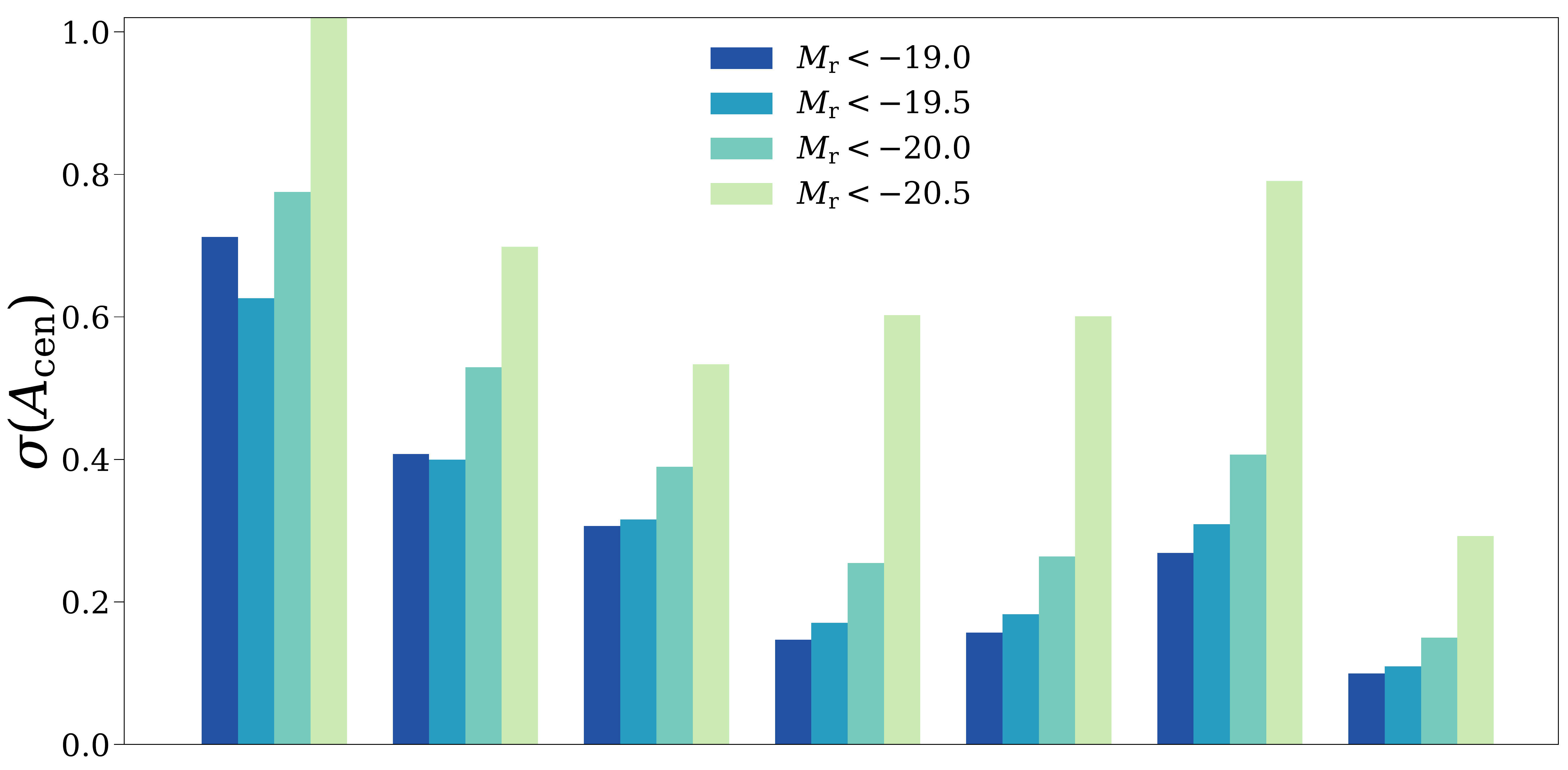} 
\end{minipage}

\begin{minipage}{0.8\textwidth}
   \includegraphics[width=1\linewidth]{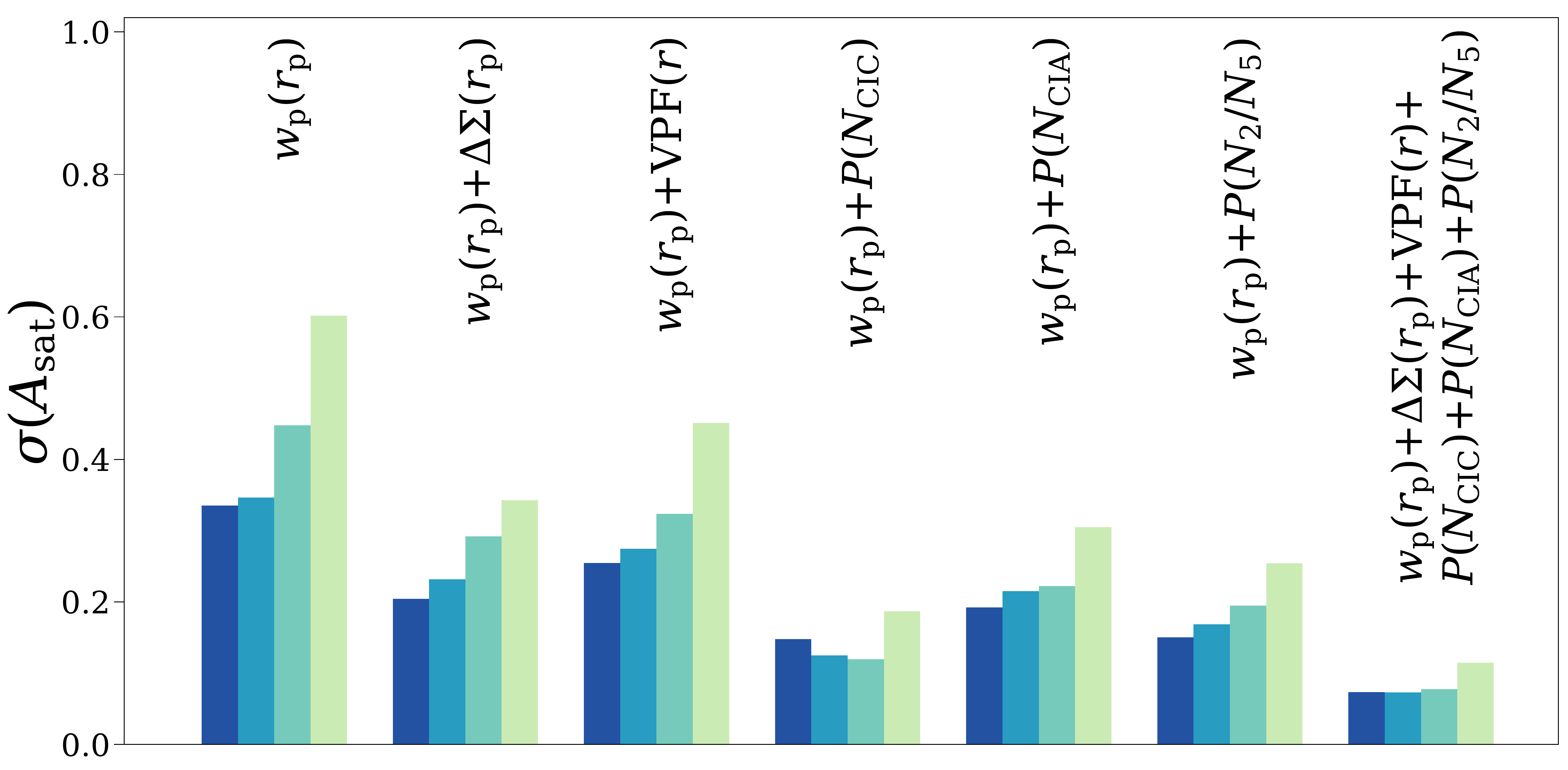}
\end{minipage}
\caption{\label{fig:Asigma}This figure shows the marginalized 1$\sigma$ constraint on $A_{\rm{cen}}$ (top panel), and $A_{\rm{sat}}$ (bottom panel), as grouped histograms. Each group of bars corresponds to a different combination of observables, and within each group, results for different luminosity thresholds are plotted in different colours, as detailed in the legend. We show $\wprp$ individually, its combination with every other observable, and the combination of all 6 of our observables. In the upper panel, the constraint from $\wprp$ for the $M_r<-20.5$ sample exceeds the range of the $y$-axis, and since $\Acen$ is restricted to the range between -1 and 1, $\Acen$ is unconstrained in this case.}
\end{figure*}

We now turn to a more detailed exploration of the
complementarity of various observables with galaxy clustering,
as quantified by $\wprp$. Figure~\ref{fig:Asigma} 
displays the constraining power of different combinations of observables on
the two galaxy assembly bias parameters of our dHOD model: 
$\Acen$, the central galaxy 
dHOD assembly bias parameter (top panel), 
and $\Asat$, the satellite galaxy dHOD assembly bias parameter (bottom panel). 
We include in Fig~\ref{fig:Asigma} constraints from $\wprp$ individually 
along with constraints from combining $\wprp$ 
with each of the other observables that we study. 
For completeness, we also show the constraints from all 
the observables combined 
as an illustration of the maximal constraining power 
that can be achieved using the complete set of observables considered in our study.  
The constraints in Fig.~\ref{fig:Asigma} are depicted as bar plots, with the bars 
grouped by combination of observables. 
Bars of different colours within 
each group correspond to different luminosity 
threshold samples, as indicated. 
Finally, the heights of the bars represent the fully marginalized 
1$\sigma$ constraints, with smaller values corresponding to tighter, 
more restrictive constraints. Similar plots for the other dHOD parameters 
can be found in Appendix~\ref{sec:app_constraint}.

Examining the bars in Fig.~\ref{fig:Asigma}, 
several general trends are apparent. Most 
prominently, similar to what we find for the $M_r<-20.0$ sample in Fig.~\ref{fig:vdB_4p}, 
for all the luminosity samples we study, when combined with $\wprp$,
$\Pncic$ is generally more effective at constraining 
$\Acen$ and $\Asat$ than the commonly-used $\Dsigma$, 
despite our assumption of infinite lensing source density.
In fact, in most cases the count statistics are the most 
effective observables to combine with clustering in an 
effort to constrain dHOD models of assembly bias. 
Furthermore, as is shown in Appendix~\ref{sec:app_constraint}, 
this statement is typical of the constraints on most of the HOD parameters, 
especially for the fainter samples.

When comparing results for galaxy samples defined by different luminosity thresholds, 
the constraints are typically tighter for the fainter samples. 
This mainly reflects the fact that brighter samples have lower number densities, resulting in higher levels of noise.
However, there are some exceptions. For example, the constraints on $\Acen$ from $\wprp$ alone are tighter for the $M_r < -19.5$ sample than for the $M_r < -19.0$ sample. In these cases, the degeneracies among different parameters depend on luminosity, such that after marginalization over all other parameters the noiser, high luminosity sample yields tighter constraints.
The unmarginalized constraints are all monotonically increasing functions of luminosity threshold, as they must be.

\subsection{Marginalized Two-Dimensional Constraints}

\begin{figure*}
\centering
  \includegraphics[width=\textwidth,trim={0 0 4cm 4cm},clip]{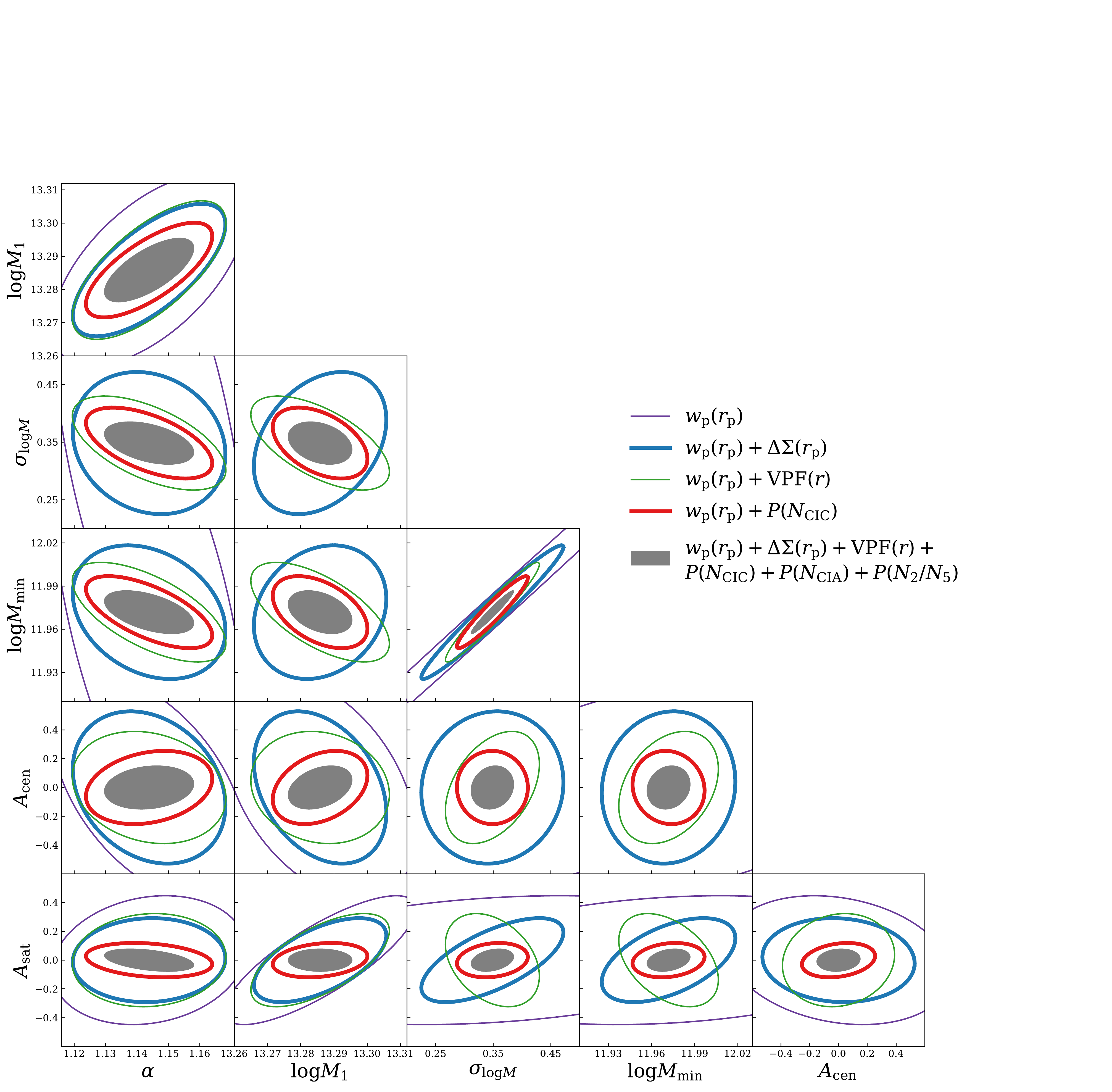}
  \caption{\label{fig:contour_wx}
  The marginalized $1\sigma$ constraint 
  contours on the dHOD parameters in each of the two-dimensional projections of 
  the dHOD parameter space for the $M_r<-20.0$ sample, from the combinations of $\wprp$ with $\Dsigma$, $\VPF$ and $\Pncic$ respectively, as well as $\wprp$ individually. Each such combination is shown in a solid line, colour coded consistently with Fig.~\ref{fig:vdB_4p}, as labelled in the legend.
  We highlight $\wprp+\Dsigma$ and $\wprp+\Pncic$ using thick solid lines. The combination of all 6 observables is shown as shaded regions in grey, to indicate the maximal constraining power in our analyses, and for cross comparison with Fig.~\ref{fig:contour_wx}. The $\wprp$ contours are not shown in some panels, because they exceed the range of the axes.
  }
\end{figure*}

In addition to fully marginalized constraints, it is interesting to 
examine parameter constraints in two-dimensional subspaces of the 
full parameter space. In Figure~\ref{fig:contour_wx}, we plot the 
marginalized 1$\sigma$ contours in each of the 2D projections of 
our 6-dimensional parameter space. As we utilize a Fisher matrix to estimate 
parameter constraints, all contours are elliptical and are centred 
around the fiducial point in the parameter space. 
Fig.~\ref{fig:contour_wx} corresponds to the $M_r<-20.0$ threshold sample, 
but the other samples have qualitatively similar features.
Different contours correspond to different observables, 
or combinations thereof, as indicated, and we have highlighted the results for 
$\wprp+\Dsigma$ and $\wprp+\Pncic$ using thicker contours. 
To avoid crowding, we use $\Pncic$ as the representative 
case for the various count statistics. For comparison, the gray shaded 
ellipse shows the constraints derived from exploiting all of our observables simultaneously.

\begin{figure*}
\centering
  \includegraphics[width=\textwidth,trim={0 0 4cm 4cm},clip]{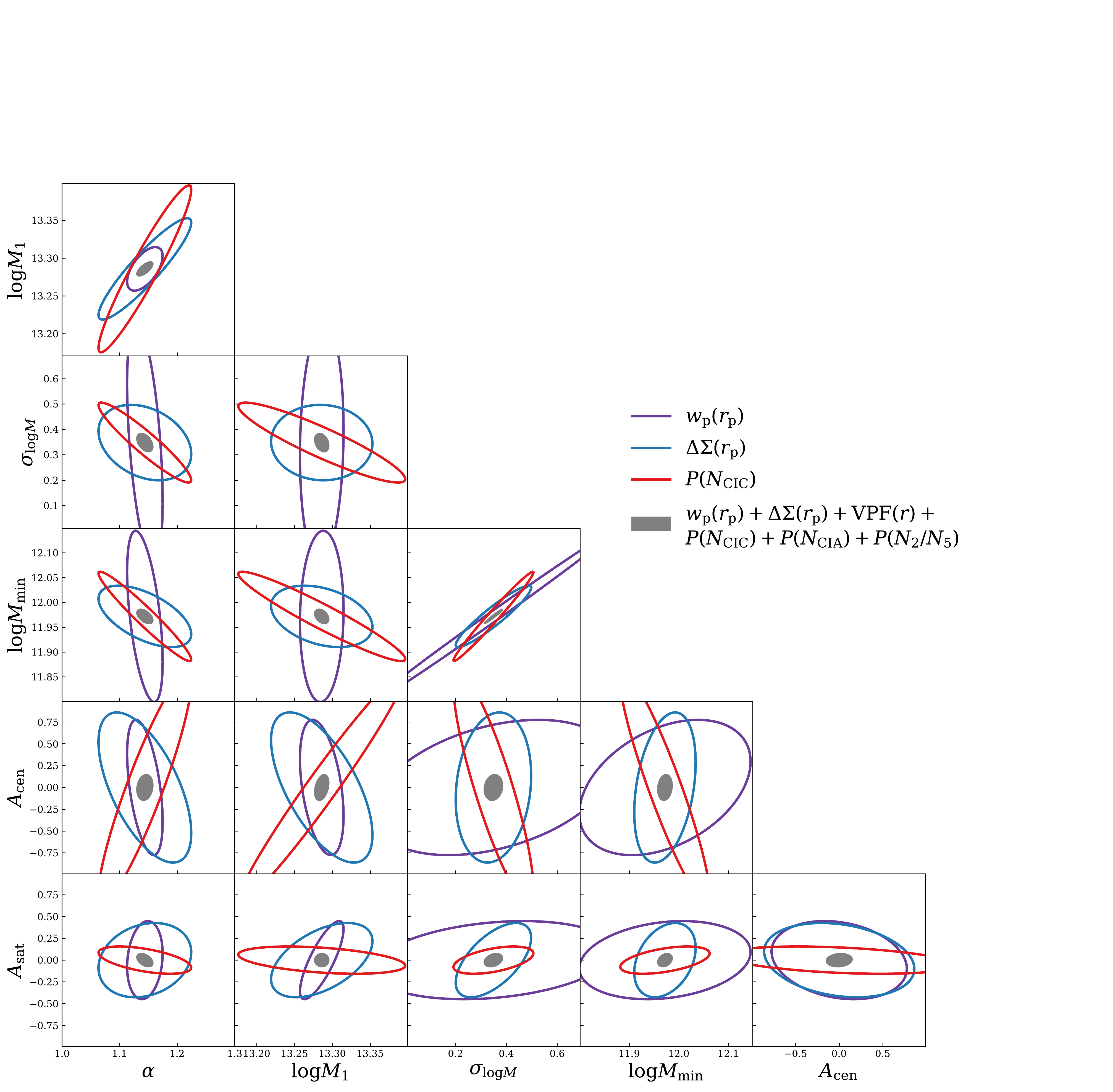}
  \caption{\label{fig:contour_x}Same as Fig. \ref{fig:contour_wx}, but for the individual observables. We only show $\wprp$, $\Dsigma$ and $\Pncic$ for clarity. The combination of all 6 observables is shown as shaded regions in grey, for cross comparison with Fig. \ref{fig:contour_wx}. 
  }
\end{figure*}
 
From Fig.~\ref{fig:contour_wx}, it is apparent that in all projections, $\wprp+\Pncic$ 
(thick red lines) is superior to the 
other combinations of observables, 
particularly the commonly-used combination of $\wprp+\Dsigma$. 
This result remains strictly true for the $M_r<-19.0$ and $M_r<-19.5$ samples, 
while for the brighter, $M_r<-20.5$ sample, 
a few parameter combinations are more tightly constrained using
parameter combinations other than $\wprp+\Pncic$. 

Figure~\ref{fig:contour_x} displays 
confidence contours for our model parameters 
constrained by $\wprp$, $\Dsigma$ and $\Pncic$ individually 
(rather than combinations of observables). 
This visualization can aid in the qualitative understanding of 
our results. Combining $\wprp$ with $\Dsigma$ yields limited 
improvement because both of these observables share 
similar degeneracy directions in multiple dimensions of the parameter space.
The combination of $\wprp$ with $\Pncic$ is 
superior because these observables have largely 
complementary degeneracy directions in the 
parameter space and combining these observables 
leads to the simultaneous breaking of multiple 
degeneracies. Notice that constraints from $\Pncic$ 
on any single parameter are not particularly restrictive; 
however, the constraints from $\Pncic$ restrict parameter values 
to exceedingly narrow degeneracy regions, 
which, in turn, leads to significant improvements in 
constraining power when combined with $\wprp$.

\subsection{Constraints on Parameters of the Standard HOD}

In the previous subsections, we focused on constraints on assembly bias 
parameters in the dHOD model and showed that counts-in-cylinders is an 
effective comlement to the galaxy projected two-point function for 
diagnosing and constraining assembly bias. However, it is also interesting 
to study constraints on the standard HOD parameters in a standard HOD model that 
does not include assembly bias.

\begin{figure*}
\centering
  \includegraphics[width=\textwidth,trim={0 0 2.2cm 4cm},clip]{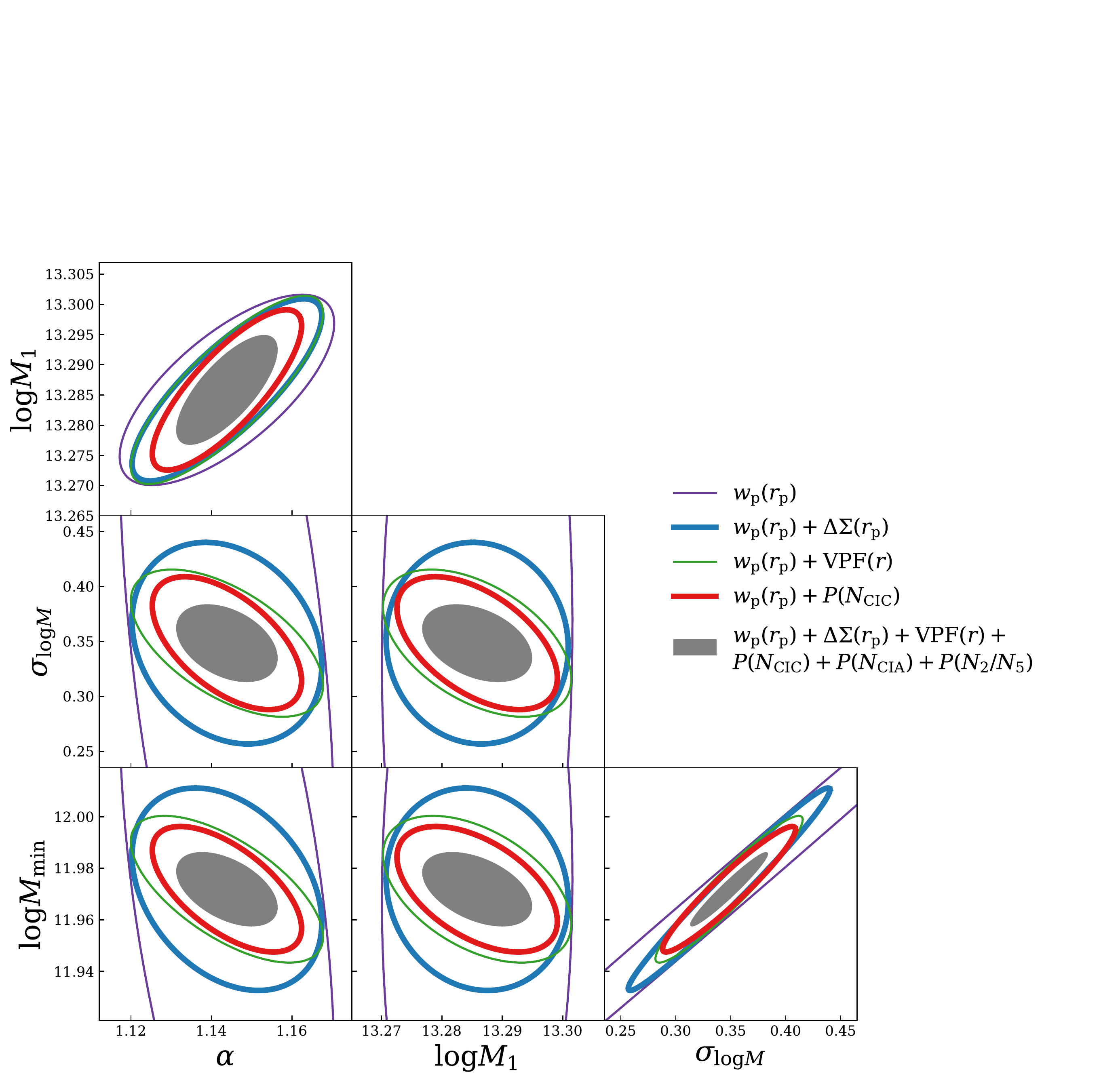}
  \caption{\label{fig:contour_wx_sHOD}Same as Fig. \ref{fig:contour_wx}, but for analysis on the standard HOD model. In this figure we show representative results for the $M_r<-20.0$ luminosity threshold sample.}
\end{figure*}

Figure~\ref{fig:contour_wx_sHOD} depicts 2D marginalized, projected constraint 
contours on the standard HOD parameters from an analysis to constrain a 
standard HOD model using the various observables that we consider. 
It is clear that the complementarity of $\Pncic$ extends to the 
parameters of the standard HOD, as the combination of $\wprp$ with 
$\Pncic$ outperforms the combination of $\wprp$ with $\Dsigma$ in all 
projections. This strongly suggests that $\Pncic$ is a favourable 
observable even in studies using standard HOD.

\subsection{Limitations and Caveats}

Our results are subject to several limitations and caveats 
that we describe in this subsection. 
First, all Fisher matrix analyses are based on a 
linear expansion of the likelihood: 
The observable--parameter relation 
is treated as a linear function of the observable on the parameter.
This linearity is not an accurate 
model of the observable--parameter relation over the 
entirety of the relevant domain of parameters for all of the 
observable--parameter combinations that we explore.  
As a result, the derivative values $\partial \mathbf{f}/\partial \mathbf{p}$, 
and therefore the constraints, depend on the choice of the 
fiducial point in the parameter space.
For the results presented above, the fiducial point is 
motivated by previous data analyses using a standard HOD, 
specifically, the study of \citet{zentner_etal16}. 
The fiducial values of the assembly bias parameters, 
$\Acen$ and $\Asat$ are set to zero.
We address this particular caveat in the following subsection, 
pointing out the dependence of our conclusions upon the 
fiducial location in the parameter space.

Second, the decorated HOD parameters $\Acen$ and $\Asat$ 
can only vary over the interval $[-1,1]$ because the 
degree of galaxy assembly bias that is possible is 
limited \citep[see][for details]{hearin_etal16}. 
The Fisher formalism assumes a multivariate Gaussian posterior 
distribution, so such hard boundaries on the parameter space 
can lead to gross violations of this assumption. 
In particular, any time that $\sigma_{\Acen}$ or 
$\sigma_{\Asat}$ approach unity (or even exceed it, 
see Appendix~\ref{sec:app_constraint}), our estimates will 
not be a reliable, quantitative estimate of the constraining 
power of the observables. However, the {\em qualitative} 
comparisons among observables should not be impacted 
by this shortcoming and the Fisher matrix will still 
give a reliable ranking of the relative utility of 
different combinations of observables.

An additional caveat to our results is associated with 
the particular model that we explore. While we phrase 
our results qualitatively in terms of constraints on 
galaxy assembly bias, it is important to realize that 
our calculations pertain only to a specific model, namely 
the dHOD with a binary split on galaxy populations. 
It is possible that our conclusions would change 
significantly if a different halo occupation model is used. Examples 
of different models might include a standard HOD 
with an augmented set of parameters or a wholly 
different model for the galaxy--halo relationship, 
such as the conditional luminosity function \citep[e.g.,][]{yang03} 
or a parameterized form of abundance matching \citep[e.g.,][]{Lehmann2017}. 
There are a limited number of models that include tunable galaxy assembly 
bias based on halo properties. Aside from the dHOD, \citet{Lehmann2017} 
parametrizes the dependence of galaxy luminosity on halo concentration 
at a fixed halo mass in the abundance matching model with an interpolation scheme, 
and \citet{tinker08b}, \citet{wibking_etal19} and \citet{mcewen_weinberg18} 
allow for local density-dependent variations of the HOD.
We limit our results to the dHOD model \new{and assembly bias due to concentration}, and relegate more 
comprehensive studies of galaxy assembly bias to future work.

Our constraints from the galaxy--galaxy weak lensing signal $\Delta\Sigma$ represent
the upper limit of information that can be gained in a real data analysis, 
because we use a weak lensing covariance that includes {\em only} sample 
variance. This is equivalent to assuming an infinite background source galaxy density 
and thus a shape-noise-free measurement of the galaxy--galaxy lensing signal. 
Since the galaxy--galaxy lensing signal presented is optimistic, 
our primary qualitative result, namely that 
the combination of $\wprp$+$\Pncic$ is superior 
to combinations that include the galaxy--galaxy lensing signal, 
will not be affected when complete galaxy--galaxy lensing covariances 
are used.

As a final caveat, we emphasize that our work uses only a single 
simulation and thus, we work in the context of a single set of 
cosmological parameters. In particular, we consider the best-fit 
Planck cosmology used as the modelling framework in the Bolshoi Planck simulation. 
This limitation is difficult to circumvent at this time due to the need for 
simulations that are both large volume (to model clustering) and high resolution 
(to measure the internal properties of haloes). The computational costs of 
such simulations prohibit simultaneous explorations of assembly bias 
and cosmology within the scope of this paper. However, important steps are 
being taken in precisely this direction \citep[e.g.,][]{zhai_etal18}.
Nonetheless, the Planck constraints on cosmological parameters 
are quite restrictive \citep{Planck2018,Planck2018_cosmoparams} 
and we do not expect modifications to the cosmological model to have 
a significant impact on our qualitative results.

\subsection{Dependence on Fiducial Parameters}

One of the caveats mentioned in the previous section is that 
Fisher analyses yield results that may depend upon the fiducial 
point in the parameter space about which the likelihood is expanded. 
Another way to say this is that the constraints depend upon the 
point in parameter space that corresponds to the {\em true} 
underlying model. In the results we presented above, we 
assumed that the true fiducial model corresponded to 
zero galaxy assembly bias ($\Acen = \Asat = 0$).

To examine how our results depend on this choice of the fiducial model,
we now repeat our analysis for two alternative assumptions for the 
fiducial values of the galaxy assembly bias parameters. 

In the first, we adopt a fiducial central galaxy assembly 
bias of $\Acen=0.5$, which is motivated by the recent analysis of galaxy clustering by \citet{zentner_etal16}, while
keeping $\Asat=0$. In this case, we find results that are 
both qualitatively and quantitatively similar to our 
main model with $\Acen = \Asat = 0$. 

We have also explored the dependence of our forecast constraints 
on the underlying amount of satellite assembly bias. To do so, 
we repeated our analyses with the fiducial satellite assembly bias 
parameter set to $\Asat = -0.6$. This value of 
$\Asat$ has several motivations. First, \citet{zentner_etal16} 
showed that clustering of galaxies in the SDSS $M_r<-19.5$ threshold 
sample is consistent with significantly 
negative values of $\Asat$. Furthermore, it is known that 
the abundance of dark matter subhaloes is anti-correlated with 
host halo concentration \citep[e.g.,][]{zentner_etal05, Mao2015, Jiang_vdB17}. Since subhaloes are believed to host satellite galaxies, this anti-correlation implies a negative
value for $\Asat$.

\begin{figure*}
\centering
\begin{minipage}{0.8\textwidth}
   \includegraphics[width=1\linewidth]{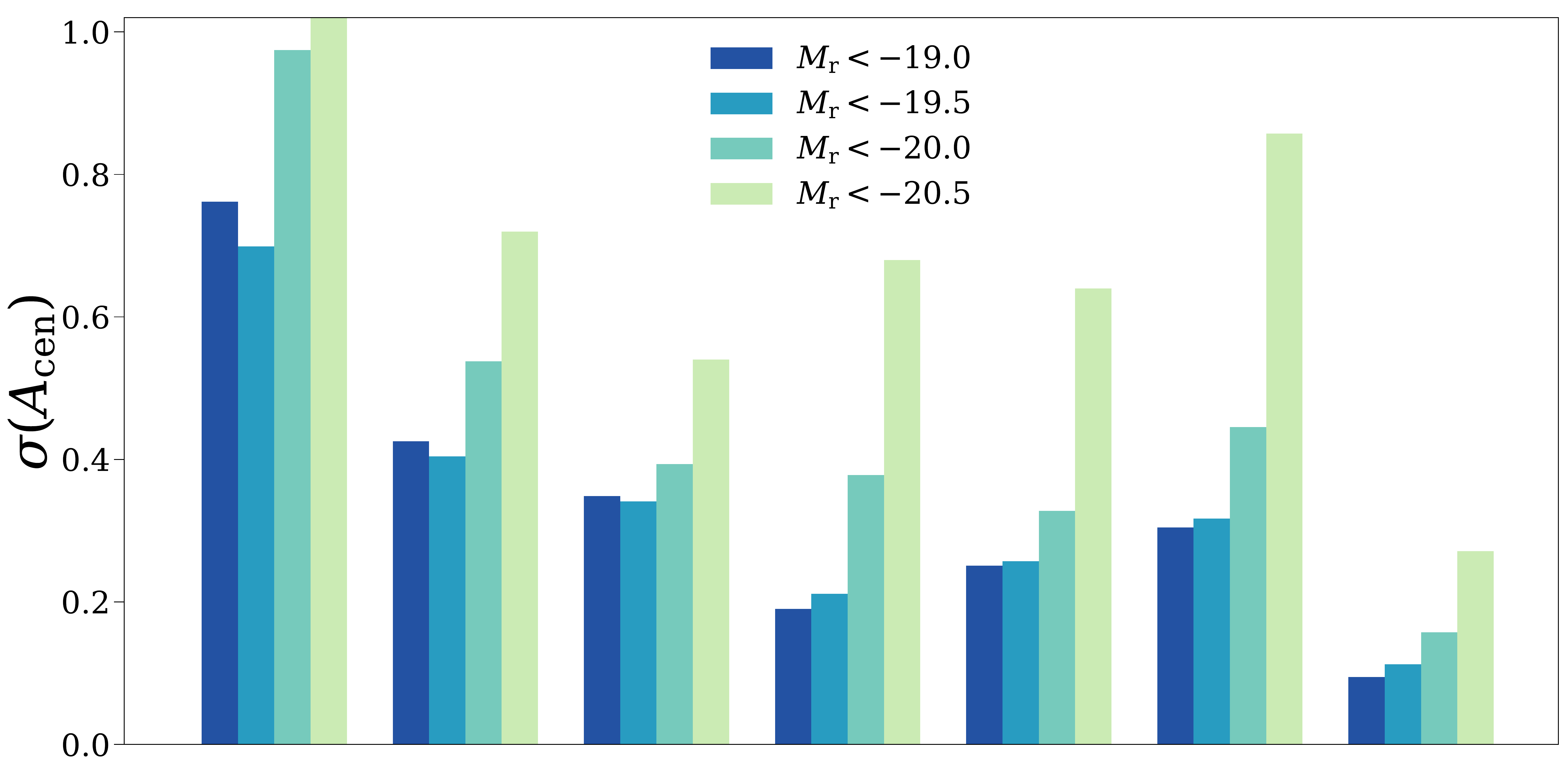}
\end{minipage}

\begin{minipage}{0.8\textwidth}
   \includegraphics[width=1\linewidth]{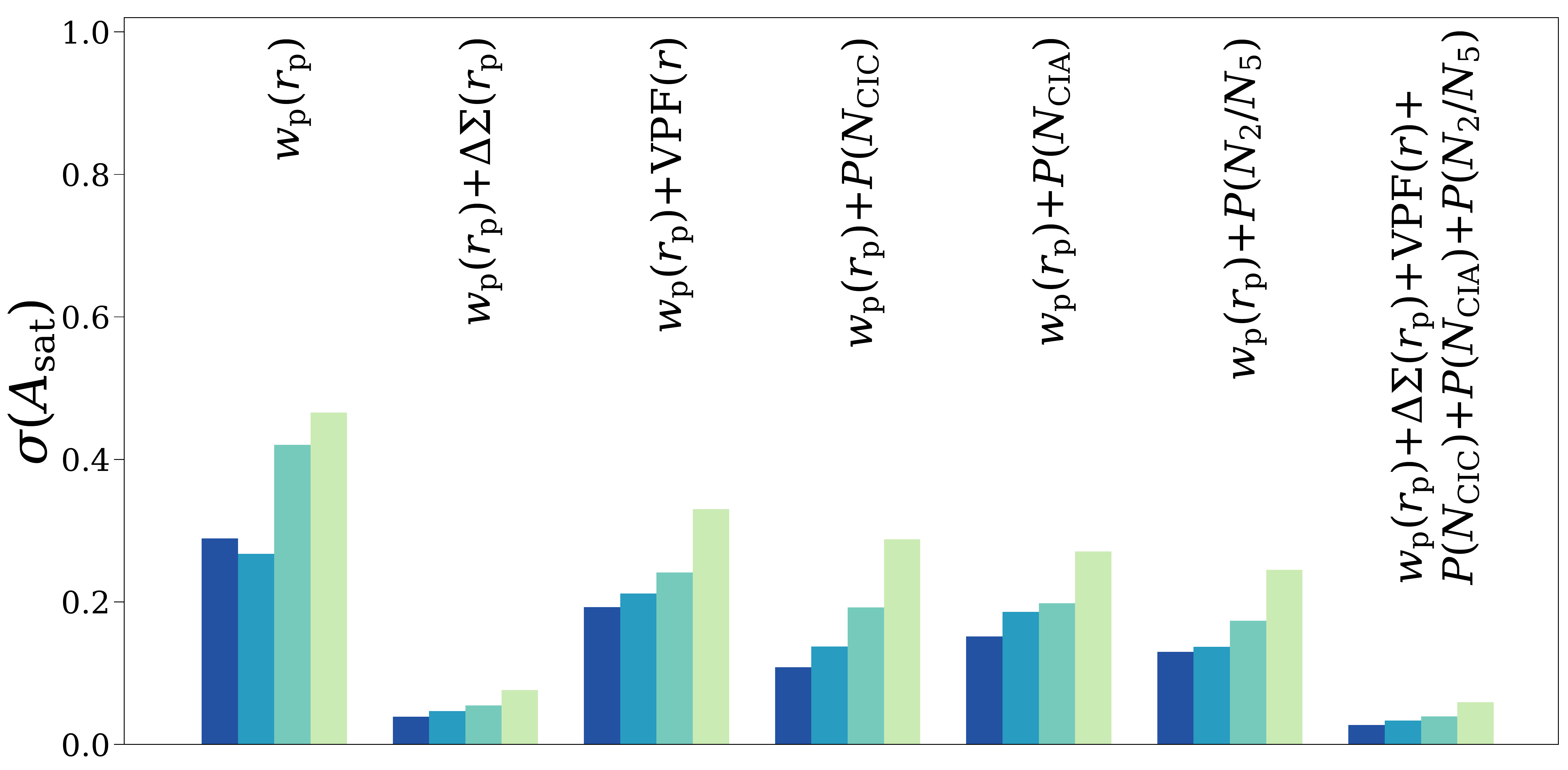}
\end{minipage}
\caption{\label{fig:Asigma_negAsat}Same as Fig.~\ref{fig:Asigma}, but with fiducial $\Asat=-0.6$.}
\end{figure*}

The results of this experiment are shown in Fig.~\ref{fig:Asigma_negAsat}. 
While the results for $\Acen$ (upper panel) are similar to the case of our 
main model with  $\Acen = \Asat = 0$ (i.e., $\Acen$ is always best constrained 
by the combination of $\wprp$ plus $\Pncic$), the results for $\Asat$ (lower panel) 
are notably different. In particular, the combination of $\wprp$ and $\Dsigma$ 
now yields the {\it tightest} constraints on $\Asat$, rather than the weakest. 
This very tight constraint stems from two things. First, 
$\wprp$ on small scales ($\lesssim 1\, \Mpch$) 
has a much stronger dependence on $\Asat$ near $\Asat=-0.6$. 
This improves constraints from $\wprp$ alone, reducing parameter 
degeneracy. Second, the observables $\wprp$ and $\Dsigma$ are 
more complementary to one another near $\Asat=-0.6$ because the 
degeneracy directions selected by the $\wprp$ constraints 
change their orientation slightly in the parameter space. 
These improvements jointly boost the constraining power of the combination 
of $\wprp$ and $\Dsigma$. However, we must note that 
several observational and theoretical factors that will 
likely impact the constraining power of very-small-scale clustering statistics
are not accounted for in our analysis. Chief among these omissions is our neglect of 
shape noise in galaxy--galaxy lensing covariances. Including shape noise is likely 
to reduce significantly the complementarity of $\Dsigma$ with other probes, including 
$\wprp$. Modelling uncertainties, such as the choice of the radial distributions 
of satellite galaxies will also reduce the constraining power of observables on 
small-scales with similar result. 
Hence, the tight constraints on $\Asat$ we observed here 
are likely to be quite optimistic.

To summarize, the results that we have shown so far point 
toward a clear conclusion: Count-based galaxy clustering 
statistics, such as counts-in-cells distributions ($\Pncic$), 
can be instrumental in constraining galaxy assembly bias, 
and are particularly powerful in constraining the 
galaxy assembly bias of central galaxies.
For reference, we tabulate the forecast constraints on all dHOD parameters 
from all of the observable combinations we study, 
and for all five luminosity threshold samples in 
Appendix~\ref{sec:app_constraint}.

\section{Discussion and Conclusions}
\label{sec:discussion}

Constraining galaxy assembly bias is important to 
the study of the connection between galaxies and haloes 
and for extracting the maximum possible information on 
both galaxy evolution and cosmology from survey data.
Numerous studies use galaxy clustering 
to constrain either the galaxy--halo connection or 
cosmology or both 
\citep[e.g.,][]{hawkins_etal03,vdBosch03a,yang_etal04,
seljak05,seljak06,cooray06,vdBosch07,zheng07,moster10,zehavi_etal11,anderson_etal12,guo_etal12, guo_zehavi_zheng12,leauthaud_etal12,cacciato_etal13,Reddick2013,guo_etal14,
reid_etal14,coupon_etal15,more_etal15,saito_etal16,lange2018,cowley_etal18,sinha_etal18, xuhj_2018}.
Several of these works combine clustering with 
either weak galaxy--galaxy lensing measurements 
or with measurements of redshift space distortions
in order to constrain the galaxy--halo connection 
and/or cosmology, and this use of complementary 
variables is becoming increasingly common. 
However, these results may suffer from 
systematic bias when galaxy assembly bias is
not properly included in the model \citep[e.g.,][]{zentner_etal14}.
To date, there are only a small number of 
studies constraining assembly bias with galaxy clustering data 
\citep{Lehmann2017,zentner_etal16,vakili_2019}.

We have studied the ability of various galaxy clustering 
statistics to constrain assembly bias in an effort to determine 
which combination(s) of observables are most informative. 
In particular, we have estimated the relative constraining power 
of several spatial galaxy clustering statistics to constrain 
the assembly bias parameters of the decorated halo occupation distribution (dHOD). 
In this first study of its kind, we have restricted our attention to 
statistics that are not particularly sensitive to galaxy 
peculiar velocities (however, we do work in redshift 
space so our results are not completely immune to 
peculiar velocities). We have chosen to do this because 
including peculiar velocities (such as redshift space distortions) 
opens up many distinct modeling questions. Exploring redshift space 
distortions will be the subject of future work.

In general, we have found that the combination of $\wprp$ with 
the counts-in-cylinders distribution 
($\Pncic$) is the most effective 
combination of two observables for constraining assembly bias within 
the dHOD modeling framework. This combination outperforms 
the commonly-used combination of $\wprp$ and $\Dsigma$ in constraining central 
galaxy assembly bias by a factor of $\sim 2$ for lower 
luminosity samples, and to a slightly lesser extent in 
constraining satellite galaxy assembly bias. The primary 
reason for this is that $\wprp$ and $\Dsigma$ share 
roughly common degeneracy directions in the dHOD parameter space, 
while the combination $\wprp$ and $\Pncic$ 
is much more complementary. This implies that the 
combination of $\Dsigma$+$\Pncic$ is nearly as 
good as $\wprp$+$\Pncic$, which we have confirmed. 
The complementarity between $\Dsigma$ and $\Pncic$ is not unexpected.
The lensing signal traces the matter density contrast 
around galaxies while the counts-in-cylinders statistics 
probe the galaxy distribution profiles in approximately 
the same regions, and are therefore expected 
to complement each other in constraining cosmology 
as well as the connection between galaxies and 
the matter field \citep{gruen_des_lensing_cic2018, friedrich_des_model_lensing_cic2018}. 

We have shown that the count statistics are also effective
in constraining the standard HOD parameters, 
independent of whether the actual halo occupation 
statistic are affected by galaxy assembly bias or not.
In addition, we have tested different fiducial models 
and find that when strong satellite assembly bias is present, 
the combination of $\wprp+\Dsigma$ actually 
provides the tightest constraints on $\Asat$ 
among all sets of observables studied here.
Therefore we caution that the preferred statistics 
may depend on the true, underlying relationship between 
galaxies and dark matter haloes, and the degree to which 
galaxy assembly bias is realized in nature.

Our results complement recent work studying the information that can be gained from higher-order statistics beyond the two-point function. In particular, in \citet{yuan_etal17,yuan_etal18} it was shown that the three-point function in the squeezed limit contains significant additional constraining power on HOD parameters that is complementary to $\wprp.$ This is consistent with our findings in the sense that the complementarity of counts statistics to $\wprp$ derives precisely from the extraction of information in higher-order $n$-point moments of the density field. Considering the results in \citet{yuan_etal18} together with our findings, statistics beyond two-point clustering and lensing should be seriously considered in future analyses of large-scale structure data that utilize models of the galaxy-halo connection.

In conclusion, our results strongly suggest that galaxy assembly bias may 
be significantly better constrained and/or better understood by employing simple 
counts statistics as measured from forthcoming and present-day data sets. 
There are numerous forthcoming data sets with which can be used to inform assembly bias 
and/or whose interpretation may be challenged by small levels of assembly 
bias. These include large redshift surveys, such as may be carried out by 
DESI \citep{levi2013_desi} or WFIRST \citep{WFIRST2015}. As we have already mentioned, 
interesting follow up work includes an 
exploration of velocity statistics, 
such as redshift space distortions, 
in redshift surveys. 
While our work relates specifically to redshift surveys, it would 
be interesting to explore possible avenues for studying assembly bias 
within photometric surveys, such as the DES 
\citep{DES2005,DES2018} and LSST \citep{LSST2009}. 
It is our aim to study and deploy these statistics to constrain 
the galaxy--halo connection and to encourage others to do the same.

\section*{Acknowledgements}

We thank Brian Batell, Arthur Kosowsky, Rachel Mandelbaum, Dazhuang Mao, Jeffrey Newman, Troy Raen, Li-Cheng Tsai, and Rongpu Zhou for useful discussions. 
KW, ARZ and ASV are supported by the US National Science Foundation (NSF) 
through grant AST 1517563 and by the Pittsburgh Particle Physics 
Astrophysics and Cosmology Center (PITT PACC).  ASV is additionally supported by the Argonne Leadership Computing Facility, which is a DOE Office of Science User Facility supported under Contract DE-AC02-06CH11357.
YYM is supported by PITT PACC through the 
Samuel P.\ Langley Postdoctoral Fellowship. FvdB and JUL are supported by the US National Science Foundation (NSF) through grant AST 1516962. FvdB received additional support from the Klaus Tschira foundation, and from the National Aeronautics and Space Administration through Grant No.\ 17-ATP17-0028 issued as part of the Astrophysics Theory Program. CMS is supported by
NSF through grant DMS 1521786.
APH is supported under the DOE contract DE-AC02-06CH11357. DC is supported by a McWilliams Postdoctoral Fellowship.

\new{The authors gratefully acknowledge the Gauss Centre for Supercomputing e.V. (www.gauss-centre.eu) and the Partnership for Advanced Supercomputing in Europe (PRACE, www.prace-ri.eu) for funding the MultiDark simulation project by providing computing time on the GCS Supercomputer SuperMUC at Leibniz Supercomputing Centre (LRZ, www.lrz.de).
The Bolshoi simulations have been performed within the Bolshoi project of the University of California High-Performance AstroComputing Center (UC-HiPACC) and were run at the NASA Ames Research Center.}

This research made use of Python, along with many community-developed or maintained software packages, including
IPython \citep{ipython},
Jupyter (\https{jupyter.org}),
Matplotlib \citep{matplotlib},
NumPy \citep{numpy},
and SciPy \citep{scipy}.
This research made use of NASA's Astrophysics Data System for bibliographic information.

\bibliographystyle{mnras}
\bibliography{main,software}

\appendix

\section{Derivative Fitting}
\label{sec:app_deriv}

In this appendix, we describe our approach to obtain the 
derivative matrix $\partial \mathbf{f}/\partial \mathbf{p}$.

We begin from the fiducial parameter set specified in Table~\ref{tab:fidp}, 
and add perturbations to each dHOD parameter 
(except $\Mzero$, which we do not vary in our analysis) in turn, 
keeping all other parameters fixed, 
to examine the dependence of observables on each individual parameter 
(i.e., the set of $\mathbf{f}$--$\mathbf{p}$ relations).
The process is non-trivial because we construct our observables from 
mock catalogues based on the population of a simulation of finite size 
with galaxies drawn from the (d)HOD. The fact that we construct 
our observables using mock catalogues has the advantage of accuracy 
compared to analytic approximation methods, but it also has the 
unfortunate consequence that individual observables can exhibit non-negligible 
stochasticity from one mock catalogue realization to another. 
In order to mitigate the impact of this variability, 
rather than taking $\Delta \mathbf{f}/\Delta \mathbf{p}$ at a single perturbed 
point, we consider a series of perturbed models in the 
neighbourhood of the fiducial value, along each dimension of the parameter space. We then fit our set of $\mathbf{f}$--$\mathbf{p}$ relations for a slope, as an estimate of the partial derivative in the neighbourhood 
of the fiducial point. We will elaborate on these procedures in the remainder of this appendix.

As we have discussed in Section~\ref{sec:decorated}, 
in the \texttt{Halotools} implementation of the (d)HOD model, 
the galaxy occupation of each halo is randomly drawn from the 
probability distribution function determined by the properties of the halo.
The mean occupation varies with the (d)HOD parameters, 
whereas any particular realization of the mock galaxy catalogue 
is also dependent on the sequence of random numbers used in this process. 
To generate the mock catalogues that are suitable for Fisher analysis, we use a fixed random seed for each random variable in the (d)HOD model. The values of random seeds are set independently of the (d)HOD parameter values. In this fashion, the number of galaxies in each halo will always have the same random deviate each time the halo is populated from the underyling (d)HOD model, regardless of the parameter values.
Had we not implemented this, 
the $\mathbf{f}$--$\mathbf{p}$ relations we measure would be severely 
impacted by the significantly greater stochasticity introduced by the random process. 
It would then be required to construct a very large number of mock catalogues to 
marginalize over this stochasticity. By employing a common seed, 
the differences between these catalogues become primarily attributable 
to parameter differences and not statistical fluctuations from catalogue 
to catalogue that could be induced by finite sampling of the (d)HOD models.
This exercise minimizes the variation of the observables attributable to stochasticity; 
however, $\mathbf{f}$ is generally not a smooth function of $\mathbf{p}$, 
due to the intricate nature of the dependence of the observables on the galaxy distribution.

\begin{figure}
\centering
  \includegraphics[width=0.49\textwidth]{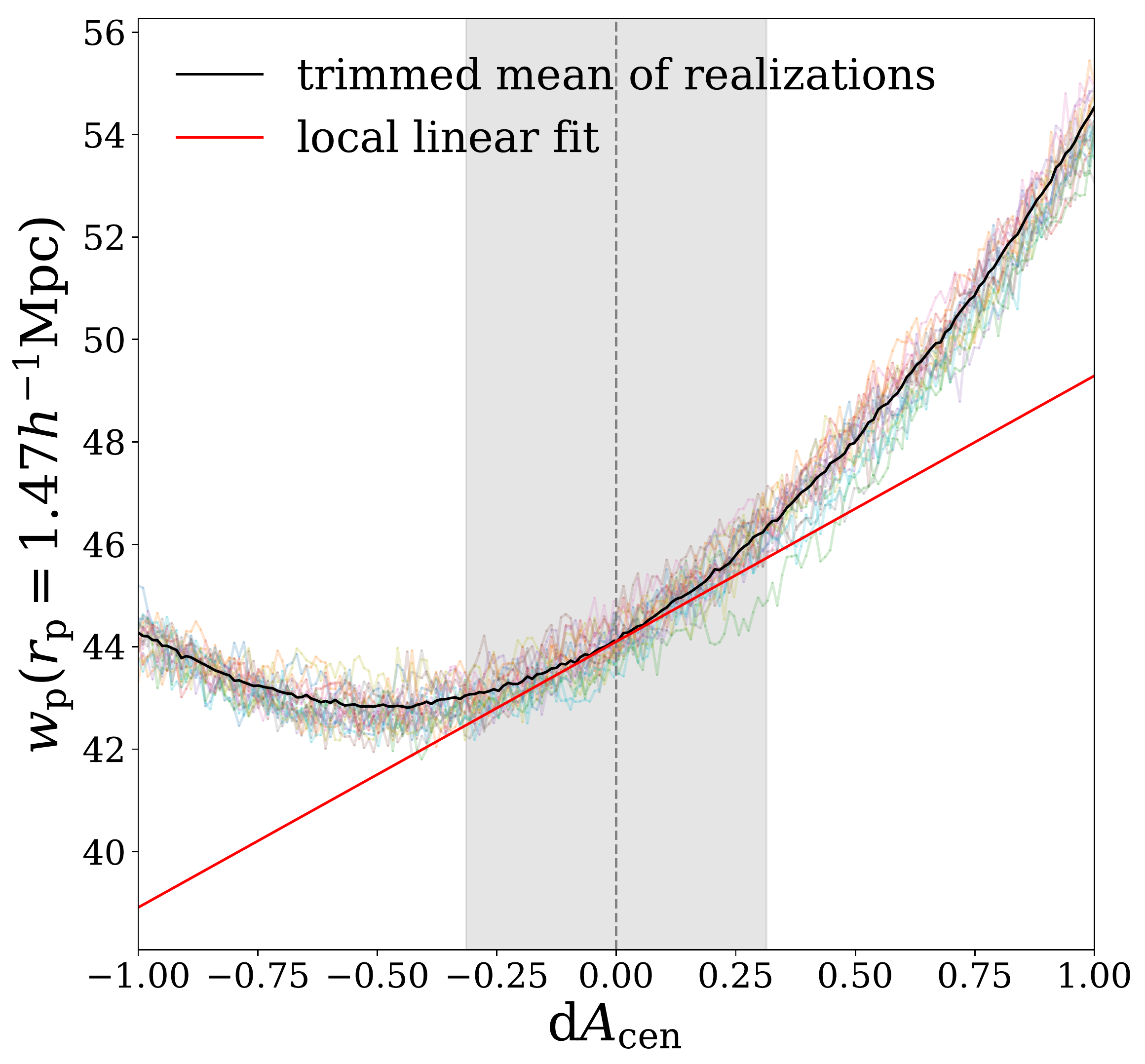}
  \caption{\label{fig:deriv_fit}In this figure, we show an example of our $\mathbf{f}$--$\mathbf{p}$ relations and its fitted derivative. In this example, we study the observable $w_{\mathrm{p}}(\rp=1.74\Mpch)$ against the perturbation in $\Acen$. Each thin coloured line is the dependence of the observable on the parameter, obtained with a different random number seed. The solid black line shows the trimmed mean of these fixed-seed $\mathbf{f}$-$\mathbf{p}$ data points. The fiducial parameter is marked by the vertical grey dashed line (in this case at $\Acen=0$, so that $\mathrm{d}\Acen=\Acen$). The smoothing scale adopted is shown as a vertical grey band, and the solid red line is the local linear fit.)}
\end{figure}

Determining the $\mathbf{f}$--$\mathbf{p}$ relations from a single random 
number seed is not sufficient because the $\mathbf{f}$--$\mathbf{p}$ relation has a 
small dependence upon the random number seed. Consequently, 
the values of the resulting derivatives will vary slightly with different random seeds.
For this reason, we repeat this entire process for a large number (or order $\sim 100$) of different 
random number seeds and take the trimmed mean of $\mathbf{f}$, averaging only the central $68\%$ values. 
In this manner, we construct smooth $\mathbf{f}$--$\mathbf{p}$ relations that 
do not dependent on the choice of the random seed. 
In the example shown in Fig.~\ref{fig:deriv_fit}, 
we assess the relation between the projected two-point correlation function 
$w_{\mathrm{p}}$ at $\rp=1.74\Mpch$ and the central galaxy assembly bias parameter $\Acen$. 
Thin coloured lines correspond to different random number seeds, and for clarity we only
plot results for a subset of all random seeds used.
The thicker black line shows the trimmed mean from the central 68\% of 
the fixed-seed $\mathbf{f}$--$\mathbf{p}$ data points, which serves as 
our estimate of the observable--parameter relation. This procedure provides us with 
a set of $\mathbf{f}$--$\mathbf{p}$ relations, one for each observable--parameter pair, 
that we fit as described below.

We use the \texttt{R} \citep{R_language} 
package \texttt{locfit} \citep{locfit_R} to fit local linear 
derivatives to the $\mathbf{f}$--$\mathbf{p}$ relation that we have obtained. We choose the 
degree of local polynomials to be two, which captures the shape of the curve 
without excessive overfitting. The \texttt{locfit} package provides 5 commonly 
used weighting kernels. We have confirmed that different kernels yield similar results. 
We present results obtained with the default \texttt{tricube} kernel.
Because some of the $\mathbf{f}$--$\mathbf{p}$ relations are strongly non-linear, the smoothing scale for the local fit needs to be chosen with care. We choose the smoothing scale of fitting following 
the principle that the range of the parameter
considered for the local derivative fit should be comparable with the posterior 1$\sigma$ 
constraint for each observable combination respectively. Qualitatively, this is motivated 
by the fact that the 1$\sigma$ constraints defines what it means to be in the 
``neighbourhood'' of the fiducial point. To this criterion, 
we add two additional restrictions:
\begin{enumerate}
    \item For each $\mathbf{f}$-$\mathbf{p}$ relation, we obtain the optimal smoothing parameter from generalized cross validation (GCV), using the \texttt{loess.as} function in the \texttt{R} package \texttt{fANCOVA} \citep{fANCOVA_R}. We use this smoothing scale as a lower limit, in order to avoid fitting numerical noise.
    \item We apply another lower limit of smoothing scale for each parameter in each threshold, based on the physical interpretation of $\mathbf{f}$-$\mathbf{p}$ relations, as some of the unphysical effects from the mock are not recognized by GCV. We have tested that our conclusions are not sensitive to this choice within a reasonable range.
\end{enumerate}
 In the example of Fig.~\ref{fig:deriv_fit}, we fit a local linear derivative to the relation at the fiducial parameter set, marked by the vertical grey dashed line, i.e., $[\partial w_{\mathrm{p}}(\rp=1.74\Mpch)/\partial \Acen]|_{\Acen=0}$. The smoothing scale is shown as a grey band that is symmetric around the fiducial parameter. The local linear fit is shown by the solid red line. The fitted slopes are the derivatives we use in our forecasts.

\section{Forecast Constraints on HOD Parameters}
\label{sec:app_constraint}

In this appendix, we present a comprehensive list of our results for constraints on dHOD parameters.

\begin{figure*}
\centering
\begin{minipage}{0.8\textwidth}
   \includegraphics[width=0.99\linewidth]{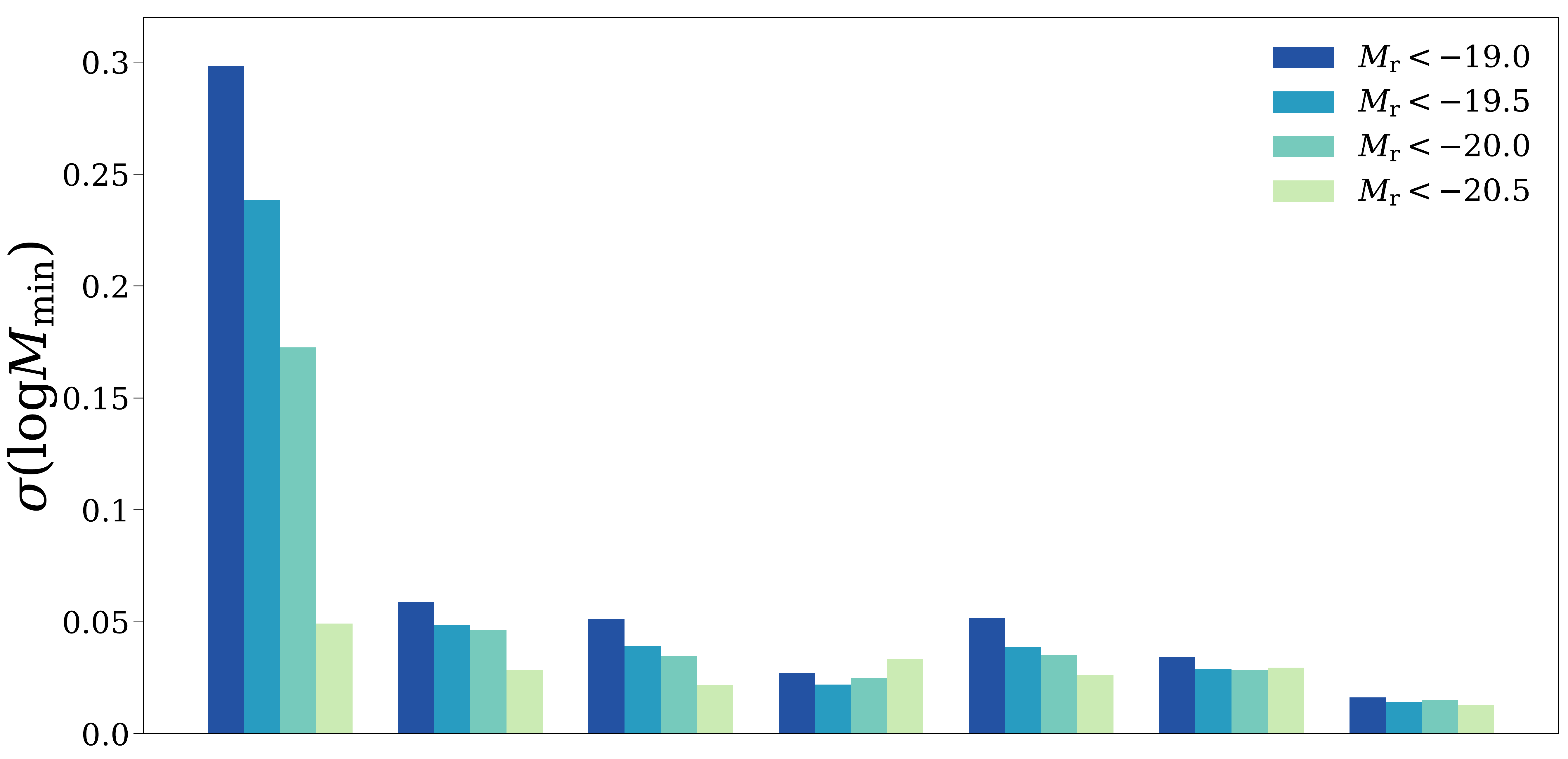} 
\end{minipage}

\begin{minipage}{0.8\textwidth}
   \includegraphics[width=0.99\linewidth]{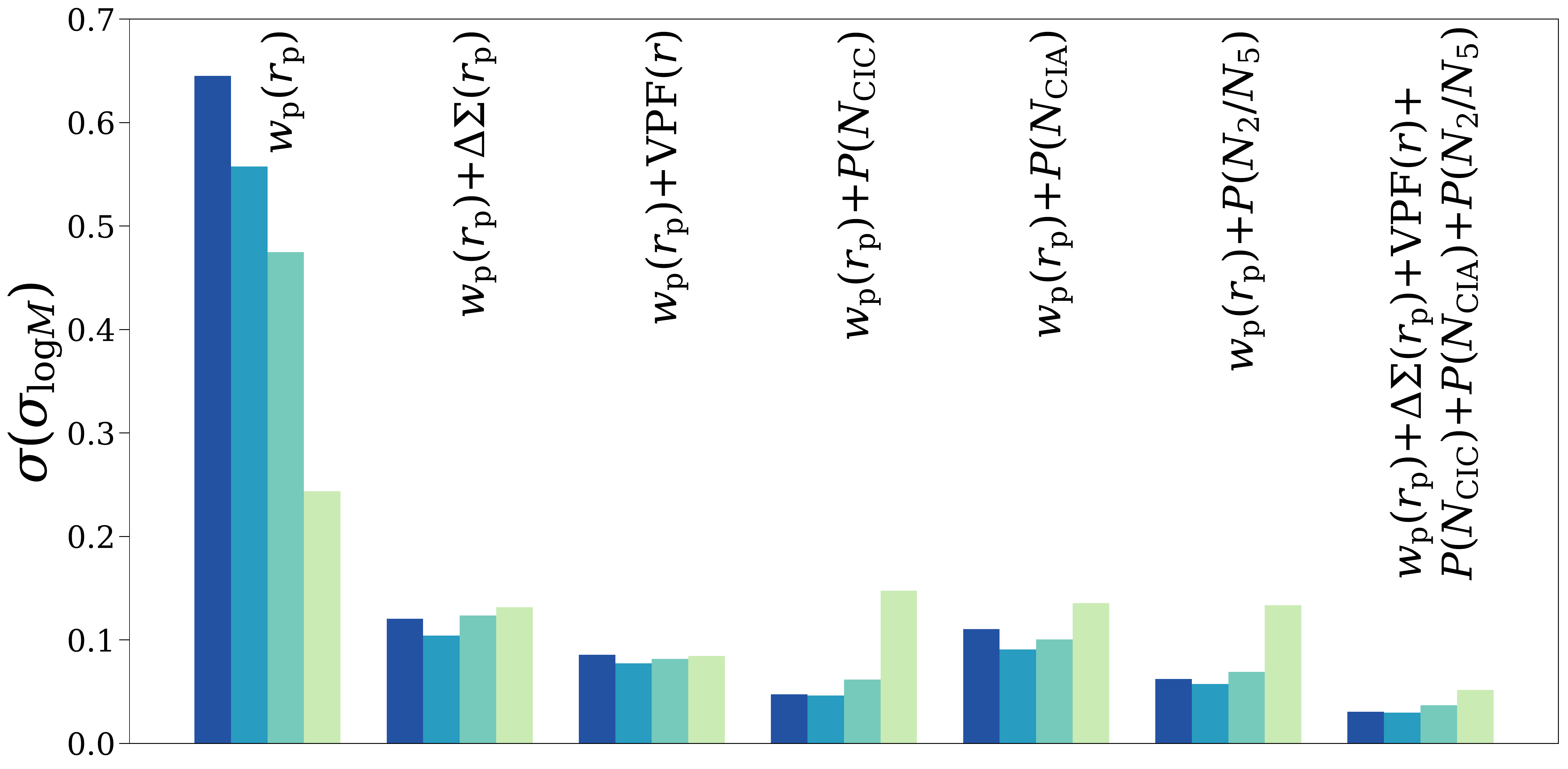}
\end{minipage}
\caption{\label{fig:Mmin_slogM}The same as Figure \ref{fig:Asigma}, but for the parameters $\alpha$ and $\log \Mone$.}
\end{figure*}

\begin{figure*}
\begin{minipage}{0.8\textwidth}
   \includegraphics[width=0.99\linewidth]{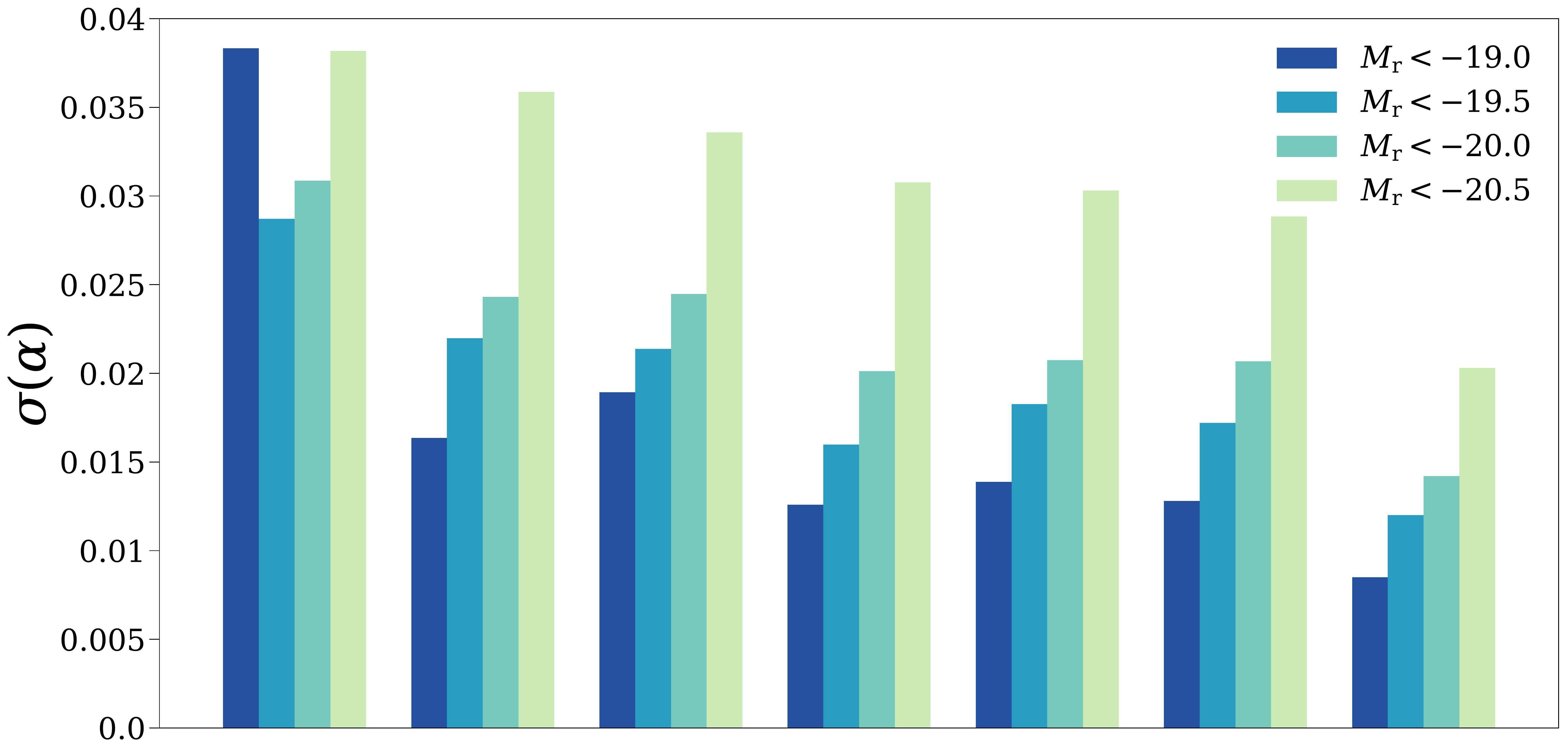}
\end{minipage}

\begin{minipage}{0.8\textwidth}
   \includegraphics[width=0.99\linewidth]{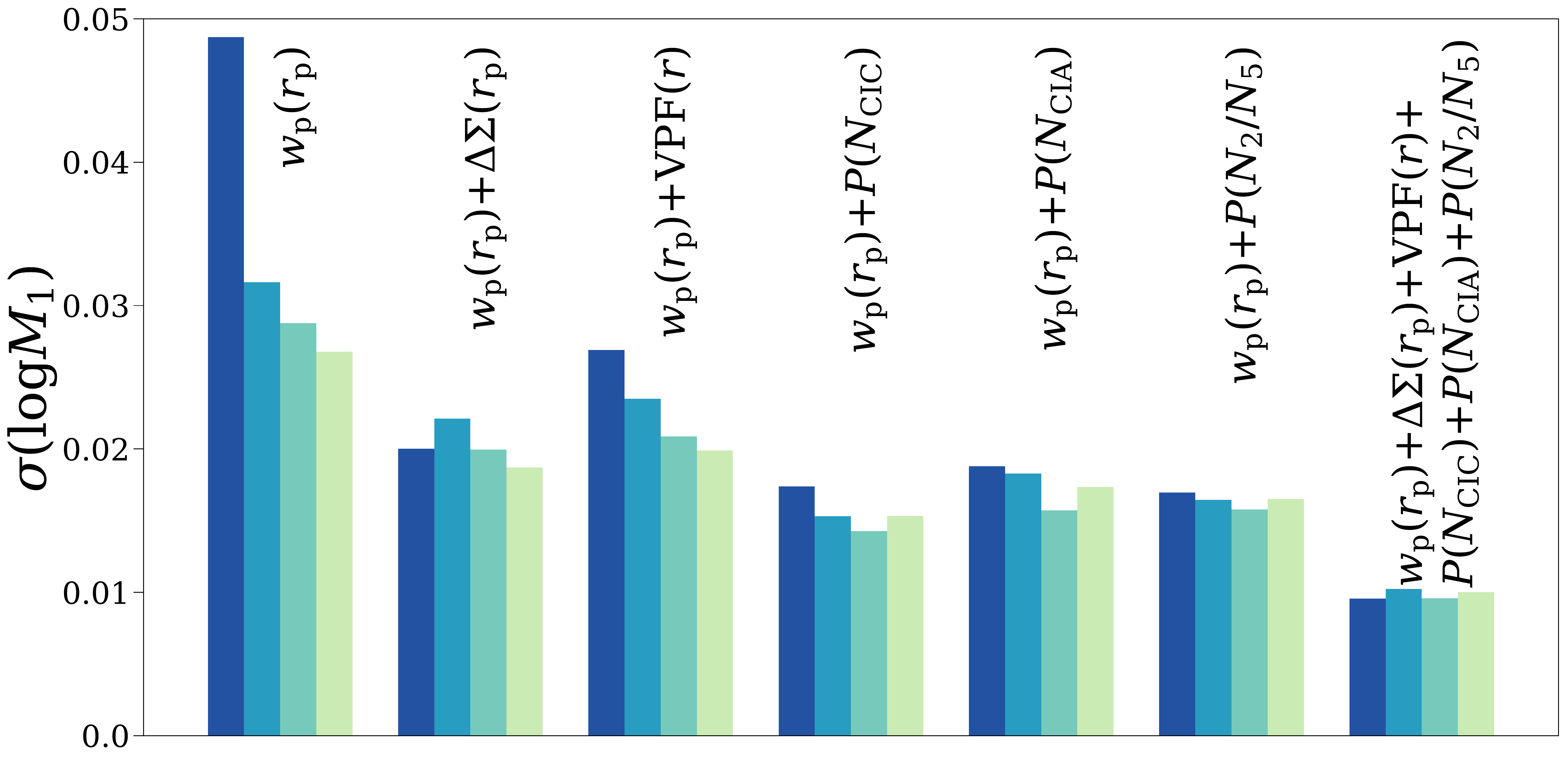}
\end{minipage}

\caption{\label{fig:a_M1}The same as Figure \ref{fig:Asigma}, 
but for the parameters $\log \Mmin$ and $\slogM$.}
\end{figure*}

In Fig.~\ref{fig:Mmin_slogM} and Fig.~\ref{fig:a_M1}, we show bar plots of constraints on the decorated HOD parameters besides $\Acen$ and $\Asat$ (shown in Figure \ref{fig:Asigma}), excluding $\log \Mzero$ which is poorly constrained. These include $\alpha$, $\log \Mone$, $\log \Mmin$, and $\slogM$, which are also the original parameters of the standard HOD.

We also list the posterior constraint values in Tables~\ref{tab:sigma190}, \ref{tab:sigma195}, \ref{tab:sigma200}, \ref{tab:sigma205}, and \ref{tab:sigma210}, for the 6 parameters that we allow to vary and for all 5 of the luminosity samples we study. The constraints we include are from all the individual observables, all the possible combinations of two observables, and the combination of all 6 observables.

\begin{table*}
	\centering
	\caption{$M_r<-19.0$}
	\label{tab:sigma190}
	\begin{tabular}{|c|c|c|c|c|c|c|}
		\hline\hline
	 Observable combination & $\sigma(\log\Mmin)$ & $\sigma(\slogM)$ & $\sigma(\alpha)$ & $\sigma(\log \Mone)$ & $\sigma(\Acen)$ & $\sigma(\Asat)$\\
		\hline
$\wprp$   & 0.30 & 0.645 & 0.038 & 0.05 & 0.71 & 0.34 \\
$\Dsigma$ & 0.10 & 0.157 & 0.066 & 0.09 & 0.79 & 0.32 \\
$\VPF$    & 0.22 & 0.129 & 0.689 & 0.37 & 0.63 & 3.99 \\
$\Pncic$  & 0.18 & 0.179 & 0.091 & 0.18 & 1.34 & 0.20 \\
$\Pncia$  & 0.06 & 0.151 & 0.049 & 0.07 & 0.46 & 0.36 \\
$\Pratio$ & 0.05 & 0.073 & 0.040 & 0.06 & 0.57 & 0.21 \\
$\wprp+\Dsigma$   & 0.06 & 0.121 & 0.016 & 0.02 & 0.41 & 0.20 \\
$\wprp+\VPF$      & 0.05 & 0.086 & 0.019 & 0.03 & 0.31 & 0.25 \\
$\wprp+\Pncic$    & 0.03 & 0.048 & 0.013 & 0.02 & 0.15 & 0.15 \\
$\wprp+\Pncia$    & 0.05 & 0.111 & 0.014 & 0.02 & 0.16 & 0.19 \\
$\wprp+\Pratio$   & 0.03 & 0.062 & 0.013 & 0.02 & 0.27 & 0.15 \\
$\Dsigma+\VPF$    & 0.06 & 0.084 & 0.036 & 0.04 & 0.31 & 0.19 \\
$\Dsigma+\Pncic$  & 0.04 & 0.054 & 0.022 & 0.03 & 0.19 & 0.12 \\
$\Dsigma+\Pncia$  & 0.05 & 0.082 & 0.030 & 0.03 & 0.17 & 0.19 \\
$\Dsigma+\Pratio$ & 0.04 & 0.064 & 0.022 & 0.03 & 0.29 & 0.14 \\
$\VPF+\Pncic$     & 0.05 & 0.057 & 0.030 & 0.06 & 0.43 & 0.19 \\
$\VPF+\Pncia$     & 0.05 & 0.061 & 0.046 & 0.04 & 0.23 & 0.35 \\
$\VPF+\Pratio$    & 0.03 & 0.050 & 0.030 & 0.04 & 0.38 & 0.20 \\
$\Pncic+\Pncia$   & 0.04 & 0.055 & 0.017 & 0.03 & 0.24 & 0.14 \\
$\Pncic+\Pratio$  & 0.04 & 0.054 & 0.019 & 0.03 & 0.23 & 0.13 \\
$\Pncia+\Pratio$  & 0.03 & 0.056 & 0.023 & 0.03 & 0.23 & 0.17 \\
$\wprp+\Dsigma+\VPF+\Pncic+\Pncia+\Pratio$ & 0.02 & 0.031 & 0.009 & 0.01 & 0.10 & 0.07 \\
		\hline\hline
	\end{tabular}
\end{table*}

\begin{table*}
	\centering
	\caption{$M_r<-19.5$}
	\label{tab:sigma195}
	\begin{tabular}{|c|c|c|c|c|c|c|}
		\hline\hline
	 Observable combination & $\sigma(\log\Mmin)$ & $\sigma(\slogM)$ & $\sigma(\alpha)$ & $\sigma(\log \Mone)$ & $\sigma(\Acen)$ & $\sigma(\Asat)$\\
		\hline
$\wprp$   & 0.24 & 0.558 & 0.029 & 0.03 & 0.63 & 0.35 \\
$\Dsigma$ & 0.07 & 0.138 & 0.080 & 0.08 & 0.76 & 0.33 \\
$\VPF$    & 0.13 & 0.179 & 0.657 & 0.30 & 0.62 & 3.10 \\
$\Pncic$  & 0.09 & 0.119 & 0.088 & 0.14 & 1.06 & 0.16 \\
$\Pncia$  & 0.05 & 0.141 & 0.075 & 0.11 & 0.72 & 0.37 \\
$\Pratio$ & 0.04 & 0.076 & 0.052 & 0.05 & 0.58 & 0.28 \\
$\wprp+\Dsigma$   & 0.05 & 0.104 & 0.022 & 0.02 & 0.40 & 0.23 \\
$\wprp+\VPF$      & 0.04 & 0.077 & 0.021 & 0.02 & 0.32 & 0.27 \\
$\wprp+\Pncic$    & 0.02 & 0.046 & 0.016 & 0.02 & 0.17 & 0.12 \\
$\wprp+\Pncia$    & 0.04 & 0.091 & 0.018 & 0.02 & 0.18 & 0.22 \\
$\wprp+\Pratio$   & 0.03 & 0.057 & 0.017 & 0.02 & 0.31 & 0.17 \\
$\Dsigma+\VPF$    & 0.05 & 0.079 & 0.047 & 0.04 & 0.30 & 0.21 \\
$\Dsigma+\Pncic$  & 0.03 & 0.050 & 0.026 & 0.03 & 0.20 & 0.12 \\
$\Dsigma+\Pncia$  & 0.04 & 0.071 & 0.038 & 0.03 & 0.21 & 0.21 \\
$\Dsigma+\Pratio$ & 0.03 & 0.060 & 0.028 & 0.03 & 0.31 & 0.16 \\
$\VPF+\Pncic$     & 0.04 & 0.052 & 0.039 & 0.06 & 0.44 & 0.14 \\
$\VPF+\Pncia$     & 0.04 & 0.058 & 0.062 & 0.04 & 0.26 & 0.35 \\
$\VPF+\Pratio$    & 0.03 & 0.052 & 0.038 & 0.04 & 0.34 & 0.27 \\
$\Pncic+\Pncia$   & 0.03 & 0.052 & 0.023 & 0.03 & 0.27 & 0.12 \\
$\Pncic+\Pratio$  & 0.03 & 0.052 & 0.025 & 0.03 & 0.26 & 0.13 \\
$\Pncia+\Pratio$  & 0.03 & 0.054 & 0.032 & 0.03 & 0.26 & 0.22 \\
$\wprp+\Dsigma+\VPF+\Pncic+\Pncia+\Pratio$ & 0.01 & 0.030 & 0.012 & 0.01 & 0.11 & 0.07 \\
		\hline\hline
	\end{tabular}
\end{table*}

\begin{table*}
	\centering
	\caption{$M_r<-20.0$}
	\label{tab:sigma200}
	\begin{tabular}{|c|c|c|c|c|c|c|}
		\hline\hline
	 Observable combination & $\sigma(\log\Mmin)$ & $\sigma(\slogM)$ & $\sigma(\alpha)$ & $\sigma(\log \Mone)$ & $\sigma(\Acen)$ & $\sigma(\Asat)$\\
		\hline
$\wprp$   & 0.17 & 0.475 & 0.031 & 0.03 & 0.78 & 0.45 \\
$\Dsigma$ & 0.06 & 0.149 & 0.081 & 0.07 & 0.86 & 0.43 \\
$\VPF$    & 0.10 & 0.155 & 0.783 & 0.21 & 0.56 & 2.73 \\
$\Pncic$  & 0.09 & 0.157 & 0.081 & 0.11 & 1.32 & 0.16 \\
$\Pncia$  & 0.04 & 0.137 & 0.088 & 0.11 & 0.91 & 0.33 \\
$\Pratio$ & 0.05 & 0.112 & 0.064 & 0.05 & 0.80 & 0.32 \\
$\wprp+\Dsigma$   & 0.05 & 0.123 & 0.024 & 0.02 & 0.53 & 0.29 \\
$\wprp+\VPF$      & 0.03 & 0.082 & 0.024 & 0.02 & 0.39 & 0.32 \\
$\wprp+\Pncic$    & 0.03 & 0.062 & 0.020 & 0.01 & 0.25 & 0.12 \\
$\wprp+\Pncia$    & 0.04 & 0.100 & 0.021 & 0.02 & 0.26 & 0.22 \\
$\wprp+\Pratio$   & 0.03 & 0.069 & 0.021 & 0.02 & 0.41 & 0.19 \\
$\Dsigma+\VPF$    & 0.04 & 0.083 & 0.054 & 0.04 & 0.39 & 0.25 \\
$\Dsigma+\Pncic$  & 0.03 & 0.065 & 0.031 & 0.02 & 0.30 & 0.12 \\
$\Dsigma+\Pncia$  & 0.03 & 0.076 & 0.042 & 0.03 & 0.30 & 0.22 \\
$\Dsigma+\Pratio$ & 0.03 & 0.070 & 0.033 & 0.02 & 0.43 & 0.19 \\
$\VPF+\Pncic$     & 0.03 & 0.059 & 0.033 & 0.04 & 0.44 & 0.13 \\
$\VPF+\Pncia$     & 0.03 & 0.069 & 0.071 & 0.04 & 0.32 & 0.32 \\
$\VPF+\Pratio$    & 0.03 & 0.069 & 0.040 & 0.03 & 0.38 & 0.30 \\
$\Pncic+\Pncia$   & 0.03 & 0.067 & 0.027 & 0.03 & 0.38 & 0.11 \\
$\Pncic+\Pratio$  & 0.03 & 0.074 & 0.030 & 0.03 & 0.36 & 0.12 \\
$\Pncia+\Pratio$  & 0.03 & 0.073 & 0.036 & 0.03 & 0.34 & 0.24 \\
$\wprp+\Dsigma+\VPF+\Pncic+\Pncia+\Pratio$ & 0.01 & 0.037 & 0.014 & 0.01 & 0.15 & 0.08 \\
		\hline\hline
	\end{tabular}
\end{table*}

\begin{table*}
	\centering
	\caption{$M_r<-20.5$}
	\label{tab:sigma205}
	\begin{tabular}{|c|c|c|c|c|c|c|}
		\hline\hline
	 Observable combination & $\sigma(\log\Mmin)$ & $\sigma(\slogM)$ & $\sigma(\alpha)$ & $\sigma(\log \Mone)$ & $\sigma(\Acen)$ & $\sigma(\Asat)$\\
		\hline
$\wprp$   & 0.05 & 0.244 & 0.038 & 0.03 & 1.07 & 0.60 \\
$\Dsigma$ & 0.05 & 0.197 & 0.098 & 0.06 & 1.05 & 0.52 \\
$\VPF$    & 0.07 & 0.117 & 1.023 & 0.11 & 0.71 & 2.57 \\
$\Pncic$  & 0.07 & 0.282 & 0.070 & 0.06 & 1.53 & 0.25 \\
$\Pncia$  & 0.04 & 0.209 & 0.126 & 0.12 & 1.88 & 0.40 \\
$\Pratio$ & 0.04 & 0.186 & 0.060 & 0.03 & 1.42 & 0.41 \\
$\wprp+\Dsigma$   & 0.03 & 0.132 & 0.036 & 0.02 & 0.70 & 0.34 \\
$\wprp+\VPF$      & 0.02 & 0.084 & 0.034 & 0.02 & 0.53 & 0.45 \\
$\wprp+\Pncic$    & 0.03 & 0.148 & 0.031 & 0.02 & 0.60 & 0.19 \\
$\wprp+\Pncia$    & 0.03 & 0.136 & 0.030 & 0.02 & 0.60 & 0.30 \\
$\wprp+\Pratio$   & 0.03 & 0.134 & 0.029 & 0.02 & 0.79 & 0.25 \\
$\Dsigma+\VPF$    & 0.03 & 0.083 & 0.065 & 0.03 & 0.51 & 0.32 \\
$\Dsigma+\Pncic$  & 0.04 & 0.140 & 0.039 & 0.02 & 0.57 & 0.20 \\
$\Dsigma+\Pncia$  & 0.02 & 0.106 & 0.054 & 0.03 & 0.53 & 0.27 \\
$\Dsigma+\Pratio$ & 0.03 & 0.118 & 0.039 & 0.02 & 0.67 & 0.24 \\
$\VPF+\Pncic$     & 0.02 & 0.084 & 0.036 & 0.03 & 0.60 & 0.22 \\
$\VPF+\Pncia$     & 0.02 & 0.083 & 0.073 & 0.03 & 0.53 & 0.37 \\
$\VPF+\Pratio$    & 0.02 & 0.088 & 0.042 & 0.03 & 0.58 & 0.38 \\
$\Pncic+\Pncia$   & 0.03 & 0.130 & 0.034 & 0.03 & 0.78 & 0.18 \\
$\Pncic+\Pratio$  & 0.03 & 0.142 & 0.037 & 0.02 & 0.87 & 0.20 \\
$\Pncia+\Pratio$  & 0.03 & 0.125 & 0.042 & 0.03 & 0.77 & 0.27 \\
$\wprp+\Dsigma+\VPF+\Pncic+\Pncia+\Pratio$ & 0.01 & 0.052 & 0.020 & 0.01 & 0.29 & 0.11 \\
		\hline\hline
	\end{tabular}
\end{table*}

\begin{table*}
	\centering
	\caption{$M_r<-21.0$}
	\label{tab:sigma210}
	\begin{tabular}{|c|c|c|c|c|c|c|}
		\hline\hline
	 Observable combination & $\sigma(\log\Mmin)$ & $\sigma(\slogM)$ & $\sigma(\alpha)$ & $\sigma(\log \Mone)$ & $\sigma(\Acen)$ & $\sigma(\Asat)$\\
		\hline
$\wprp$   & 0.19 & 0.308 & 0.097 & 0.03 & 0.84 & 0.90 \\
$\Dsigma$ & 0.08 & 0.101 & 0.228 & 0.07 & 0.67 & 0.81 \\
$\VPF$    & 0.12 & 0.149 & 1.558 & 0.10 & 0.53 & 4.10 \\
$\Pncic$  & 0.19 & 0.273 & 0.148 & 0.11 & 2.46 & 0.36 \\
$\Pncia$  & 0.06 & 0.070 & 0.277 & 0.15 & 1.48 & 0.88 \\
$\Pratio$ & 0.05 & 0.081 & 0.091 & 0.03 & 0.80 & 0.62 \\
$\wprp+\Dsigma$   & 0.05 & 0.074 & 0.081 & 0.02 & 0.38 & 0.57 \\
$\wprp+\VPF$      & 0.05 & 0.073 & 0.084 & 0.03 & 0.41 & 0.70 \\
$\wprp+\Pncic$    & 0.06 & 0.083 & 0.065 & 0.02 & 0.56 & 0.26 \\
$\wprp+\Pncia$    & 0.04 & 0.061 & 0.076 & 0.02 & 0.40 & 0.50 \\
$\wprp+\Pratio$   & 0.04 & 0.068 & 0.063 & 0.02 & 0.49 & 0.37 \\
$\Dsigma+\VPF$    & 0.05 & 0.062 & 0.159 & 0.04 & 0.36 & 0.56 \\
$\Dsigma+\Pncic$  & 0.05 & 0.064 & 0.073 & 0.02 & 0.41 & 0.30 \\
$\Dsigma+\Pncia$  & 0.03 & 0.044 & 0.119 & 0.03 & 0.30 & 0.47 \\
$\Dsigma+\Pratio$ & 0.04 & 0.055 & 0.077 & 0.02 & 0.37 & 0.41 \\
$\VPF+\Pncic$     & 0.06 & 0.082 & 0.073 & 0.03 & 0.51 & 0.30 \\
$\VPF+\Pncia$     & 0.04 & 0.059 & 0.205 & 0.04 & 0.38 & 0.78 \\
$\VPF+\Pratio$    & 0.05 & 0.067 & 0.083 & 0.03 & 0.40 & 0.57 \\
$\Pncic+\Pncia$   & 0.04 & 0.064 & 0.062 & 0.03 & 0.61 & 0.25 \\
$\Pncic+\Pratio$  & 0.05 & 0.072 & 0.065 & 0.03 & 0.68 & 0.27 \\
$\Pncia+\Pratio$  & 0.04 & 0.060 & 0.077 & 0.03 & 0.52 & 0.42 \\
$\wprp+\Dsigma+\VPF+\Pncic+\Pncia+\Pratio$ & 0.02 & 0.030 & 0.042 & 0.01 & 0.21 & 0.18 \\
		\hline\hline
	\end{tabular}
\end{table*}

\bsp	%
\label{lastpage}
\end{document}